\documentclass[12pt]{spieman}  
\usepackage{amsmath,amsfonts,amssymb}
\usepackage{graphicx}
\usepackage{setspace}
\usepackage{tocloft}
\usepackage{enumitem}

\usepackage{hyperref}

\usepackage{parskip}

\title{Guest Editorial: Overview of the Special Issue and a Dialog on Starshades}

\author[a]{Jonathan W. Arenberg}
\author[b]{Anthony D. Harness}
\author[c]{Rebecca M. Jensen-Clem}

\affil[a]{Northrop Grumman Space Systems,  One Space Park Drive, Redondo Beach California, USA, 90278}
\affil[b]{Princeton University, Princeton NJ, USA, 08544}
\affil[c]{University of California, Santa Cruz California, USA, 95064}

\cftpagenumbersoff{figure}
\cftpagenumbersoff{table}
\begin{document}
\maketitle

\begin{abstract}
This special issue is dedicated to starshades: science, engineering, technology and programmatics. Our reasons for organizing this special issue are several fold. First as a new technology and with research accomplished in many institutions, recent results are widely scattered in the literature. As such, we see great value in co-locating many of the most recent results. This guest editorial summarizes the 19 contributed papers as the result of a special call for papers. Since this is a rapidly maturing technology, we wanted to co-locate a primer with the most current work in the field. It is hoped that this primer will provide a tutorial to the starshade concept and pathway to the literature not in this issue. In doing so, we hope to widen the starshade community in terms of engineering and scientific engagements.  This tutorial takes the form of a  dialog, where frequently asked questions are answered.
\end{abstract}


{\noindent \footnotesize\textbf{*} Corresponding author: Jonathan W. Arenberg,  \href{mailto:jon.arenberg@ngc.com}{jon.arenberg@ngc.com} }

\begin{spacing}{1}   

\section{Overview of Special Issue on Starshades}
\label{sec:overview}
This special issue is dedicated to the subject of starshades. The starshade is a technology that has seen rapid development and wide interest at many institutions. Many of the advances in this field are spread over many journals and meetings. As a result they are difficult to collect in a single location in order to get a good view of the state of starshades. By gathering in one issue many papers representing the most recent research, we  hope that this special issue is a useful overview as of 2021. The editors of this special issue are members of the NASA chartered Starshade Technology and Science Working Group (TSWG). The idea for this special issue arose at a meeting and was endorsed by this body.

\begin{figure}[!htb]
\begin{center}
\includegraphics[width=\linewidth]{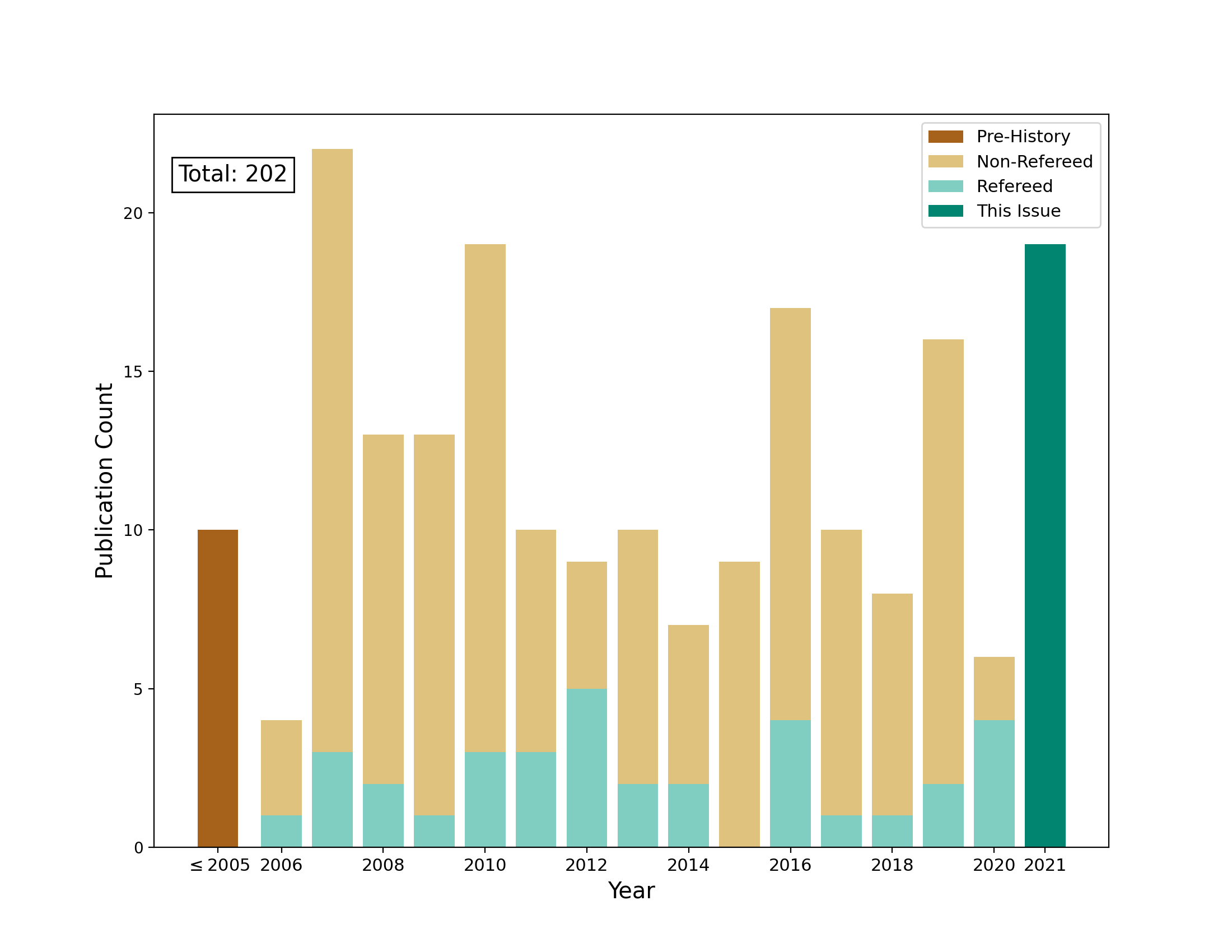}
	\caption{Published starshade papers by year. Those published before 2006 are those that are commonly recognized as being precursor papers. The data for this figure was gathered from NASA's Astrophysics Data System.}
    \label{fig:ADS}
\end{center}
\end{figure}

Figure~\ref{fig:ADS} is a plot of the number of papers about starshades on a year by year basis. The papers include both refereed and non-refereed contributions. The formal designation of the start of the starshade is 2006 when the first paper showed that a starshade can meet the basic requirements to detect extra-solar planets, especially an exo-Earth\cite{Arenberg_Cash}. The figure also shows that most of the contributions are non-refereed papers contributed at meetings and conferences. By our count the refereed contributions in this special issue constitute almost 50 percent of all of the refereed starshade literature.

As interest in developing starshade-based  missions grows, we  hope that this special issue will serve as a tutorial, providing enough of a background for potential investigators who are not familiar with starshades to have a current overview of the field in one location.  This editorial has the traditional summary of the contributed articles in Section 2. The co-editors have also created a tutorial of starshades in Section 3, which has been written in the form of a dialog between a hypothetical student and starshade ``expert''. We hope this format of ``things I am afraid to ask'' will allow those new to the field to easily ``get up to speed'' quickly and maybe have a little fun as well.

\section{Summary of Contributions}
This special issue is comprised of 19 contributed papers, representing the wide variety of research and programmatic advancement of the starshade community. The summary of papers is organized by the area of research starting with programmatics, technology (formation flying, deployment, modeling, starlight suppression), science, and concludes with mission related papers.

\subsection{Starshade Program}
{\bf Willems and Lisman} describe the goals and plans of NASA's Starshade Technology Development Activity to TRL 5 (S5). The S5 Project is the name given to the effort to close the outstanding technology gaps and raise their Technology Readiness Levels to TRL 5. The ultimate goal of this effort is to make Starshade technology ready enough when potential exoplanet missions might enter formulation. The paper describes the technology gaps and the planned efforts to close them. Attention is paid to to the consistency from which the gaps are closed.

{\bf Hu, R.~et al. -- Data Challenge} describe the Starshade Exoplanet Data Challenge: an effort to connect the science requirements of future starshade  missions with specific  performance parameters by way of synthetic images. The Data Challenge produces these synthetic images using the Starshade Imaging Simulation Toolkit for Exoplanet Reconnaissance (SISTER; {\bf Hildebrandt et al.}~) and takes into account the spatio-temporal variability of astrophysical, starshade, and telescope sources of background noise in synthetic science images. In contrast to other space-based exoplanet imaging data challenges, the Starshade Exoplanet Data Challenge includes both ``smooth'' and ``clumpy'' exozodiacal disks so that the influence of exozodiacal disks structures on exoplanet detection can be assessed. The Data Challenge is expected to conclude in September of 2021.

{\bf Turnbull et al.} report on the in-house analysis of simulated starshade observations for the Roman Exoplanet Imaging Data Challenge, a community engagement effort that examined extracting exoplanets and their orbits from simulated images for the Nancy Grace Roman Space Telescope mission. Along with data from the baselined coronagraph, the Roman Data Challenge also included simulations for the potential Starshade Rendezvous mission. In their analysis, Turnbull et al. find that the starshade's higher throughput and lower noise resulted in a 4$\times$ improvement in signal-to-noise over the Hybrid-Lyot Coronagraph (HLC). Furthermore, the combination of starshade images with radial velocity (RV) and HLC data resulted in a 2$\times$ improvement of mass and orbit constraints compared to RV+HLC data alone. This work demonstrates the enhancement in science enabled by the potential Starshade Rendezvous mission with the Roman telescope.

\subsection{Deployment and Formation Flying}
{\bf Arya et al.} report on experiments on the repeatability of the deployment of a starshade. A fundamental aspect of current starshade designs is being launched in a stowed configuration and then deploying to final shape in space. This paper describes the measurement of the two key mechanical subsystems, the inner disk and the petals. This work shows that the design currently in development can deploy with sufficient accuracy and repeatability to meet requirements.

{\bf Flinois et al.} focus on the retargeting phase of acquisition and propose an efficient guidance, navigation, and control architecture that is based solely on chemical propulsion and does not require ground tracking or interactions with the telescope during the retargeting cruise. The proposed architecture relies on commercial-off-the-shelf (COTS) accelerometers to enable actively-controlled retargeting maneuvers and a laser beacon on the starshade to initiate the acquisition phase. In their analysis, the authors use a covariance propagation framework to demonstrate this concept is feasible for the Starshade Rendezvous and HabEx mission concepts and identify the driving sources of uncertainty. The proposed retargeting architecture could help reduce the cost and complexity of the retargeting phase of a starshade mission.

{\bf Soto et al.} focus on the science phase of acquisition and present an analytical model for starshade-telescope relative dynamics to include the cost of station-keeping in observation scheduling. The dynamical model includes a deadbanding strategy to minimize maneuvers (which require an interruption in observations) and incorporates keepout angle constraints set by light from the Sun, Earth, and Moon scattering off the starshade edges. The authors present simulation results that capture the station-keeping costs as a function of target star ecliptic coordinates, time, and halo orbit phasing parameter, and find that gains can be made by optimizing the halo orbit phasing. This analytical model allows for station-keeping costs to easily be integrated into design reference mission studies.

\subsection{Starlight Suppression and Performance Modeling}
{\bf Harness et al.} report on optical verification and model validation experiments of sub-scale starshades. The experiments were conducted at the Princeton starshade testbed, an 80 m long enclosure testing 1/1000$^\mathrm{th}$ scale starshades at a flight-like Fresnel number. They demonstrate 10$^{-10}$ contrast at the starshade's inner working angle over a 10\% bandpass in the visible spectrum, with a contrast floor of 2$\times10^{-11}$. Experiments of intentionally flawed starshades validate diffraction models to better than 35\% and set an upper limit on the effect of unknowns in the diffraction models. The experiments reveal a deviation from scalar diffraction theory due to light propagating through narrow gaps between petals, but the authors provide a model that accurately captures the effect and argue this deviation is negligible at flight scales. The experiments show no optical impediments to building a starshade capable of imaging exo-Earths.

{\bf Barnett} presents an efficient algorithm for accurately solving the Fresnel diffraction equation for Fresnel numbers relevant to a starshade mission. The dynamic range in sizes over the starshade (millimeter-wide petal gaps over a diameter of 10's meters) make diffraction calculations difficult. The proposed algorithm utilizes areal quadrature and efficient nonuniform FFT algorithms to simultaneously achieve the accuracy of state-of-the-art 1D edge algorithms and the speed of FFT algorithms. The new method is shown to be in agreement with previous methods through simulations of starshades with numerically determined and hypergaussian apodization functions, but is an astonishing 10$^4\times$ faster. The author also proves the equivalence of different edge integral methods currently used in starshade modeling\cite{Cash_2011, Cady_2012}. Starshade calculations with this speed and accuracy could open the door to new strategies for designing and tolerancing a starshade mission.

\subsection{Solar Glint}
{\bf Shaklan et al.} use scatterometry results to present estimates of the brightness distribution of solar glint from uncoated metallic edges. Using custom-built scatterometers, the authors test 50 cm coupons of uncoated amorphous metal edges before and after environmental testing to produce a 2D scatter distribution function (SDF) over mission-relevant solar angles. The SDF results are incorporated into the SISTER simulation ({\bf Hildebrandt et al.}~) to estimate the solar glint brightness for the Starshade Rendezvous and HabEx mission concepts. This study finds the integrated brightness at the IWA to be $\sim$V=25 for Starshade Rendezvous and $\sim$V=27 for HabEx and find no significant change in optical performance as a result of environmental testing.

{\bf McKeithen et al.} report on the development of thin, conformal, multi-layer anti-reflection (AR) coatings to reduce solar glint brightness relative to uncoated metallic edges. Commercial finite-difference time-domain (FDTD) software was used to simulate starshade edges coated with thin-film designs that have low reflectivity at mission-relevant wavelengths. FDTD simulations show these coatings reduce not only the reflection component of scattering but also the diffraction component. The total scattering brightness was estimated to be at least an order of magnitude lower than that of uncoated metal, representing a substantial improvement over state-of-the-art uncoated edges. Experimental scatterometry measurements of a representative 50 cm long coupon were consistent with simulation results and environmental testing showed no significant change in optical performance.

\subsection{Exoplanet Detection}
{\bf Hildebrandt et al.} present the Starshade Imaging Simulation Toolkit for Exoplanet Reconnaissance (SISTER), an open-source, Matlab-based software tool for simulating images of exoplanetary systems obtained with a starshade and space telescope. SISTER models the astrophysical scene (e.g. exoplanets, exozodiacal dust, host stars, and background objects), starshade properties (e.g. the design and glint parameters), and telescope properties (e.g. the optical aberrations, bandpass, and detector noise characteristics). This code-base will be a valuable resource for planning future starshade missions.

{\bf Hu, R.~et al. -- Noise Budget} evaluate the noise budget for starshade-assisted observations with the Nancy Grace Roman Space Telescope and HabEx. The authors include a solar glint term in their noise budget that is informed by recent lab-based demonstrations of the starshade's optical performance, an effort led by NASA's Starshade Technology Development Activity to TRL5 (S5). By simulating a range of astrophysical scenes and their associated background noise contributions, the authors find that the solar glint is the dominant error term in the search of Earth-sized planets in the habitable zone  of later-type (K and M) stars, while exozodiacal light is the dominant error term for Sun-like and earlier-type stars. The authors discuss the science yield of a starshade  with Roman or HabEx under the assumption that these background terms can be calibrated to the photon-noise limit.

{\bf Hu, M.~et al.} proposes the generalized likelihood ratio test (GLRT) as a method for detecting faint exoplanets in starshade-based high contrast imaging data. The GLRT is designed to detect faint exoplanets in the presence of starshade-induced image artifacts and signal from exozodiacal dust. The authors show that the GLRT performs favorably compared to existing detection methods, on the basis of their receiver operator characteristic (ROC) curves. They also discuss the technique's astrometric and photometric precision.

{\bf Romero-Wolf et al. -- Detection Sensitivity} consider a starshade rendezvous with the Nancy Grace Roman Space Telescope. This paper calculates the sensitivity of Roman to faint exoplanets in the Starshade Rendezvous Mission architecture, focusing on three primary science cases: 1) searching for rocky planets in the habitable zones of the nearest stars and searching their atmospheres for biosignature gases; 2) detecting zodiacal dust disks around nearby sun-like stars; 3) investigating the atmospheres of known gas-giant planets at visible wavelengths.

\subsection{Observations}
{\bf Morgan et al.} compare the exo-Earth yield of the HabEx and LUVOIR mission concepts, given 1) a blind planet search, 2) partial knowledge of exoplanetary candidates and their orbits based on precursor ground-based extreme precision radial velocity observations (EPRV), and 3) perfect knowledge of all possible exoplanet systems. The authors further consider starshade only, coronagraph only, and hybrid mission architectures. The authors conclude that prior EPRV knowledge both improves the exo-Earth yield and the observing efficiency of all mission concepts, highlighting the important of EPRV precursor observations.

{\bf Romero-Wolf et al. -- Orbit Constraints} consider the science yield of the Starshade Rendezvous Mission, designed to work in tandem with the Nancy Grace Roman Space Telescope. They specifically address the ability of the Starshade Rendezvous Mission to constrain the orbits of Earth-like planets orbiting the 16 nearby stars that have been identified as the optimal targets for the mission, finding that rocky planets in the habitable zones of those stars would be detectable in at least three out of four available observing windows. They further constrain the astrometric precision and planet classification true and false positive rates.

{\bf Peretz et al. -- Performance Envelopes} use the SISTER simulation tool to examine the effects of various observational conditions on the ability to detect and spectrally characterize temperate planets in reflected light. They examine a slew of different planet types, host star types, levels of exozodiacal dust, and various starshade perturbations to set performance envelopes on a number of starshade-based missions. They find that signal-to-noise ratio requirements under-predict the  performance achieved under realistic observational conditions. They also use a statistical analysis of simulated observations to set observational completeness values for a number of accessible target stars.

{\bf Peretz et al. -- Remote Occulter Observing Constraints}  discuss the observational constraints of the Remote Occulter. The Remote Occulter is a mission for an Earth-orbiting starshade working with a ground-based telescope, designed to take advantage of the inner working angle and light collecting area of the upcoming 30-meter class ground-based telescopes. This work shows that nearly the entire sky could be observed for up to 8 hours a night. The authors investigate the sensitivity of  observational windows to the constraints and discuss the implications for future studies.

{\bf Peretz et al. -- Remote Occulter Observing Schedule} consider optimized observing scheduling with a Remote Occulter mission that includes refueling, finding that 80 exoplanet targets could be observed in the mission lifetime. The authors present a method for optimizing the science yield of such a mission, consider exposure time and observability windows, as well as time given to station-keeping and retargeting moves.

\section{A Dialog on Starshades}
As was stated earlier, in addition to the  contemporary compilation of recent work in starshades, which are of course the research articles we have summarized above, the editors would like to provide a resource to the community in the form of a review. Rather than a formal review article, the editors have reached back into astronomical history and chosen to present this review as a dialog. The use of this literary form is intended to make it easy to get a quick answer to the likely questions of one new to starshades. It is also an homage to the Galilean original \cite{Galileo} and reminds us that like the original Dialogo, this science is relatively new. The editors also thought it would be a bit of fun, and a gift from the starshade special issue editors at the end of a trying period for us all.

For our dialog, we will organize the discussion over four days, like the original. In the present case, each day will cover various aspects of the subject. Day One is for the overall basics and some history. Day Two will discuss questions regarding engineering and technology. Day Three will cover questions of science and the final day, Day Four, will address questions of a programmatic nature.   Unlike the original, our dialog involves a student of starshades, given the moniker Morgan Nemandi. Like all students of a new subject, Morgan has many questions and asks them over the course of the four day dialog.  Answering Morgan's questions is our wise and experienced starshade community acting as Morgan's tutor. As a stand-in for the collective knowledge of the starshade community and literature, we introduce Urania Sage as Morgan's guide.

\subsection{Day One: Starshade Basics}
{\bf Morgan:} Hi, my name is Morgan, and I have some questions about starshades. I hope they aren't too simple.

{\bf Urania:} Morgan, it's nice to meet you and I'm glad you are interested in starshades. They are an exciting development and should prove to be very scientifically productive. I look forward to hearing your questions, and besides, if you have a question, chances are others do too.

{\bf Morgan:} Great, let's start with the simplest question there is: very simply, what is a starshade and how does it work?

\begin{figure}[!ht]
\begin{center}
    \includegraphics[width=\linewidth]{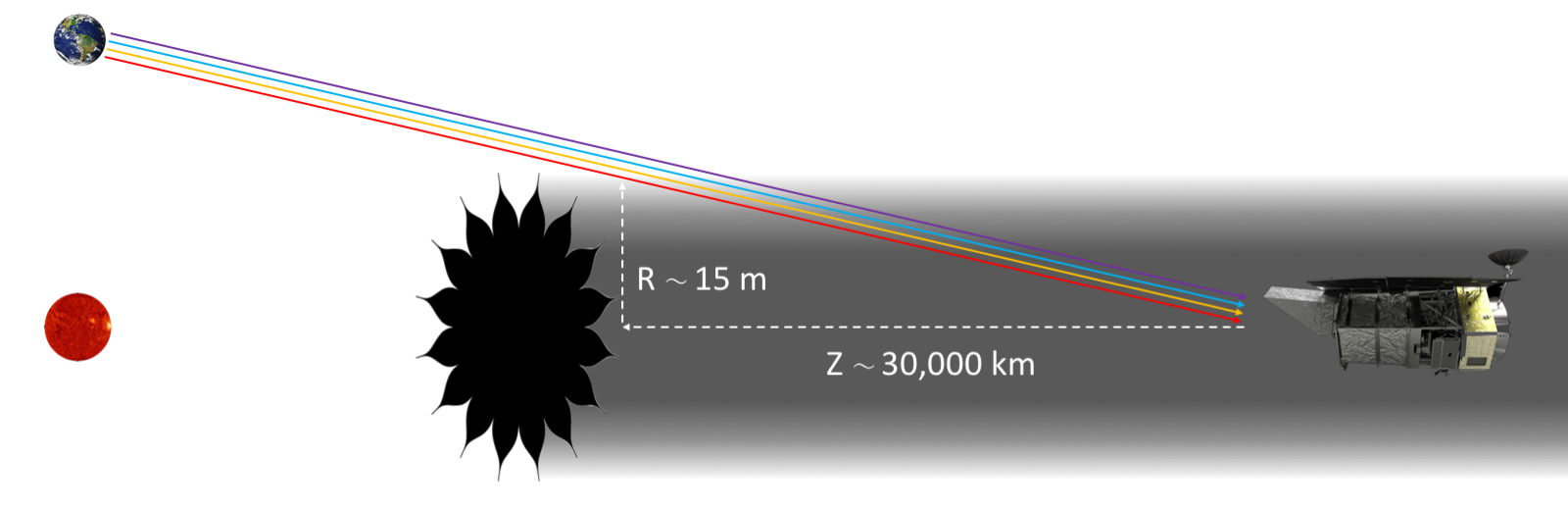}
	\caption{Cartoon of the starshade architecture. A starshade of radius $R$ located at a distance $Z$ from a telescope creates a dark shadow, which suppresses the starlight, while allowing the faint exoplanet light to enter the telescope unimpeded.}
	\label{fig:ss_arch}
\end{center}
\end{figure}

{\bf Urania:} Let's take a look at Figure \ref{fig:ss_arch}, which shows the architecture of a starshade. Morgan, please note that this picture is not at all to scale, but is drawn this way for its pedagogical purposes. From left to right, there is the host star and its planet. The starshade is next: it is a shaped optic of radius $R$, typically tens of meters. We also call it a binary optic, meaning the transmission is either complete, unity or 1, or entirely opaque, null or 0. The starshade resides on the line connecting the telescope at right and the target star at left. The spacing between the starshade and the telescope is typically tens of thousands of kilometers and is called $Z$. As shown in the picture, the starshade casts a dark shadow and the telescope sits in it, this shadow effectively turns ``off'' the host star. The light from the target planet which is off-axis relative to the star is not attenuated and is collected by the telescope.

{\bf Morgan:} So the telescope is looking over the edge of the shade, sort of like a total solar eclipse. Is it really that simple?

{\bf Urania:} Yes, conceptually it is. Implementation is a bit harder though.
  In fact, if you look carefully in the literature, you can see that external occulters have been around since the 1940's for use in solar astronomy,\cite{Evans:48} apodized or ``toothed'' apertures enter that literature in the early 1960's. \cite{Newkirk:63}

{\bf Morgan:} Speaking of ``toothed" apertures, why are starshades shaped like flowers?

\begin{figure}[!ht]
\begin{center}
    \includegraphics[width=0.5\linewidth]{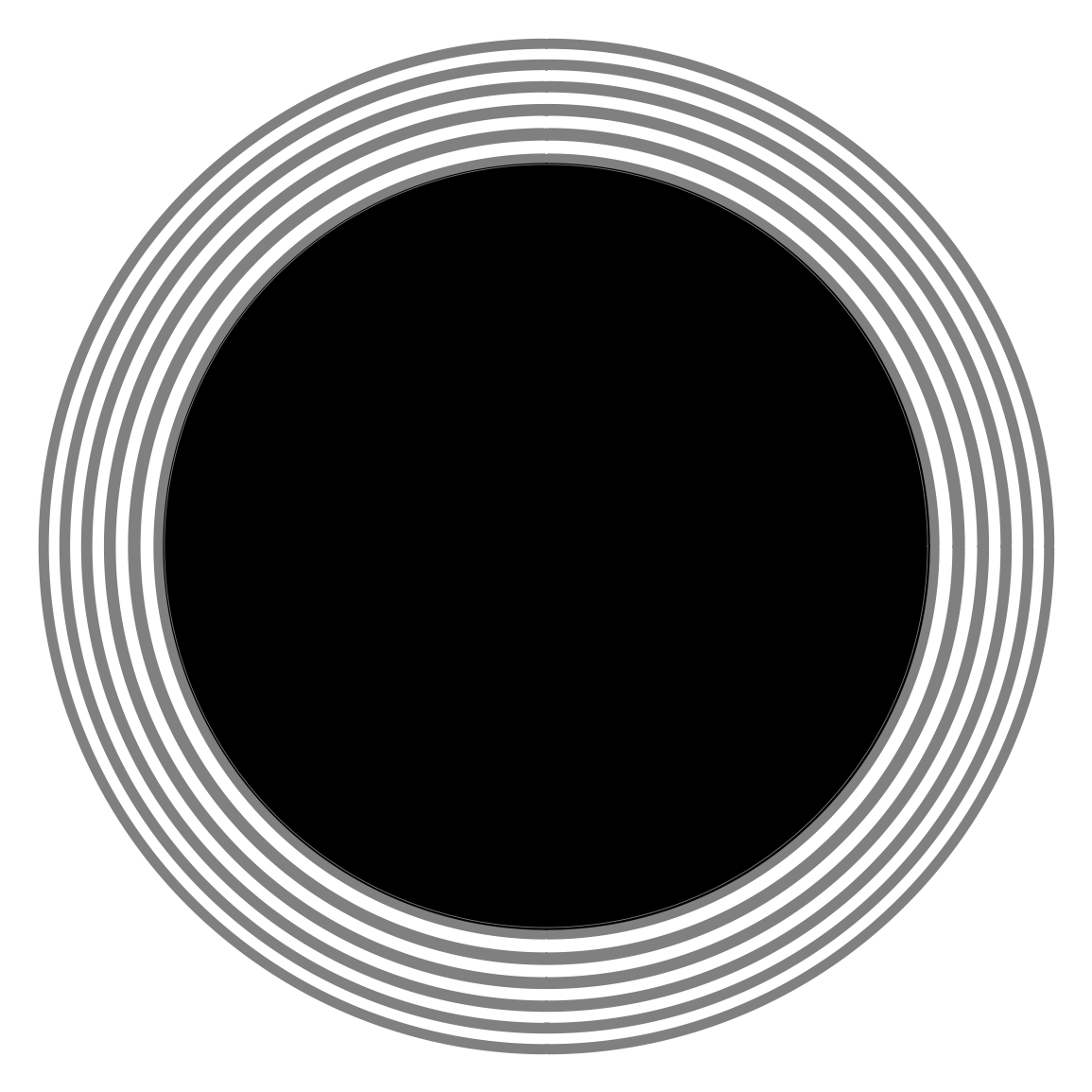}
	\caption{Circular disk, with Fresnel number of 14, in front of alternating gray and white Fresnel zones (shown out to the 26$^\text{th}$ zone).}
	\label{fig:circle_fresnel}
\end{center}
\end{figure}

{\bf Urania:} Starshades are shaped like flowers to control the diffraction of starlight around them and create a dark shadow in the image formed by the telescope. To begin to understand this fundamental aspect, let's consider Figure~\ref{fig:circle_fresnel}, which shows a simple circular disk as the black region. You also see what are called Fresnel zones, alternating gray to white, which are a mathematical construct essentially counting the difference in path length (in multiples of $\pm\lambda/2$), between the on-axis point and the radius of the zone. This difference in path length causes phase changes that determine how the electric field adds up.
\begin{figure}[!ht]
\begin{center}
    \includegraphics[width=0.5\linewidth]{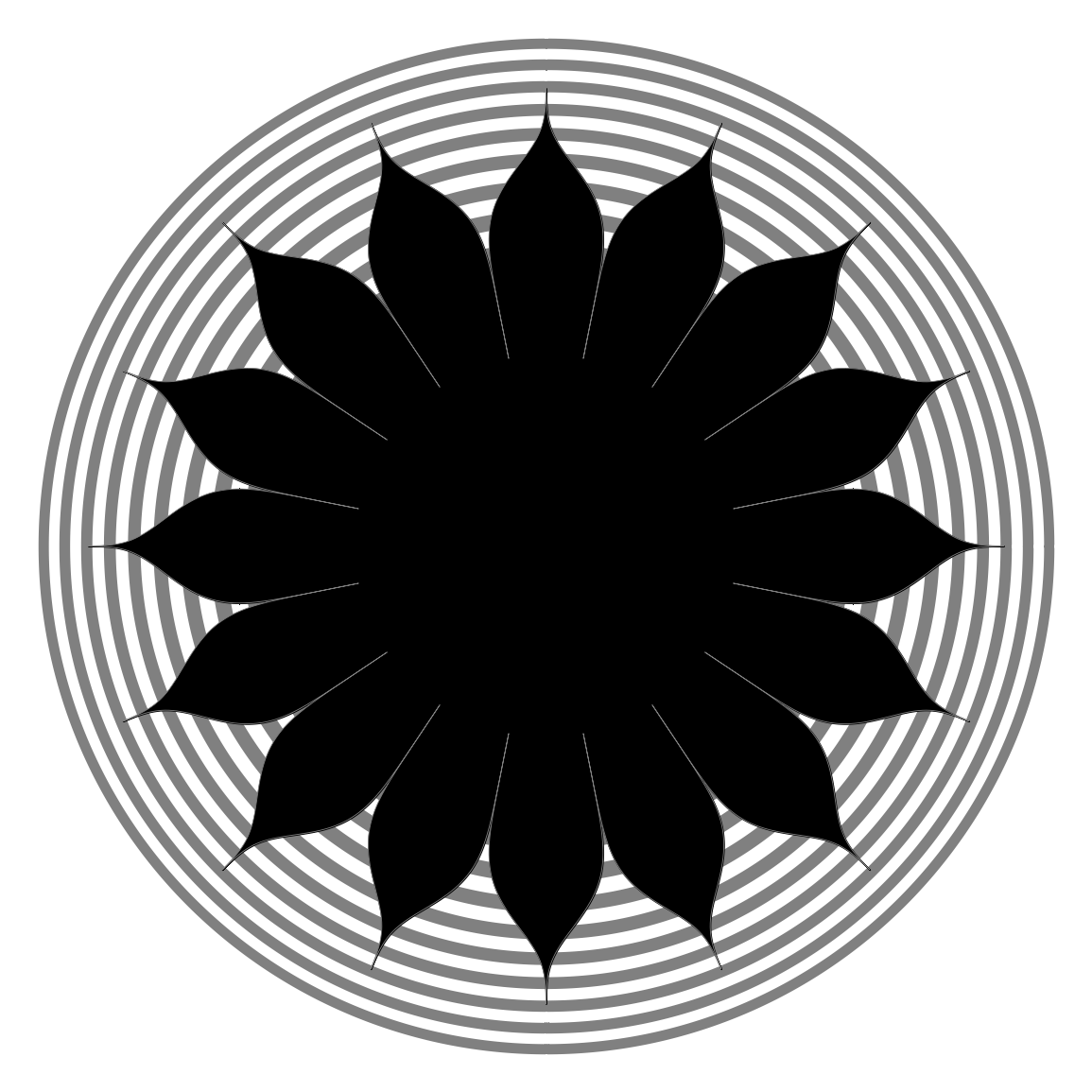}
	\caption{Sixteen petal starshade, with Fresnel number of 14 and hypergaussian design, in front of alternating gray and white Fresnel zones (shown out to the 26$^\text{th}$ zone).}
	\label{fig:starshade_fresnel}
\end{center}
\end{figure}

{\bf Morgan:} It looks to me like the distance is the same for the entire edge of the circle.
So light diffracted from the edge of the circle travels the same distance?

{\bf Urania:} Yes, exactly.

{\bf Morgan:} So everything has the same phase and adds up to the bright on-axis spot.

{\bf Urania:} Yes, Morgan, you got it!

{\bf Morgan:} So that is how we get the Spot of Arago! For any circle all the distances and phases add up  constructively\cite{Born_Wolf}.

{\bf Urania:} Correct. Since we want to ``turn off the star'', we want destructive interference, so the diffracting screen needs to vary the distance and therefore the phase, as is done in Figure~\ref{fig:starshade_fresnel}. In early literature this was called a toothed aperture.\cite{Boivin:78} Toothed apertures are well known to suppress diffraction effects.\cite{Shirley}

{\bf Morgan:} Let's see if I understand this critical piece. Our goal is to get all the light that diffracts from the starshade to the entrance aperture of the telescope to be a dark null. This will happen when light with positive phase exactly cancels the light with the opposite phase -- total destructive interference.

{\bf Urania:} Exactly! Here's another way to understand how it works. We can think of the starshade as consisting of two components: (1) a central circular disk and (2) radially tapered petals attached to the disk. The central disk serves as the occulting spot that blocks the on-axis source and creates a geometric shadow that covers the telescope. However, as we just discussed, light  will diffract around the sharp transition of the disk's edge, spoiling the darkness of the shadow. This is where the petals come in: to create a smooth reduction in the transmission of the disk to reduce the diffractive ringing caused by the sharp truncation of the electromagnetic field\cite{Spitzer_1962, Tousey_1965}. This process is called apodization and is a common problem in optical design and engineering\cite{Harris_1978, Born_Wolf}.

{\bf Morgan:} I see. This reminds me of my signal processing course where we used specific window functions to reduce the spectral leakage in Fourier transforms.

{\bf Urania:} Great observation! Yes, this is the same concept applied to the tangible world on a much larger scale. The petals are binary (either fully transparent or fully opaque) and approximate the radial apodization function that provides the critical 10 orders of magnitude of starlight suppression\cite{Marchal_1985}.

{\bf Morgan:} How does the number of petals come into this, that seems to add complexity.

{\bf Urania:} More parts is more complexity, so starshade designers want to get it right. Given a sufficient number of petals\cite{Vanderbei_2007, Cash_2011} ($n_p\gtrsim12$), the apodization function $A(\rho)$ of radius $\rho$ is well approximated by petals whose edges are defined by the polar coordinate $\theta(\rho) = A(\rho)\frac{\pi}{n_p}$. $A$ refers to the fraction of light blocked by the starshade: $A=1$ is fully opaque, $A=0$ is fully transparent. Too few petals and the approximation begins to break down.

{\bf Morgan:} It sounds like the current petal shaped starshade was not born fully formed like Minerva!

{\bf Urania:} No Morgan, but good mythology reference.

{\bf Morgan:} In my signal processing course, we used a Gaussian as our apodization function. Will any function work here? What determines the shape of the petal?

{\bf Urania:} A simple Gaussian would not work for our application, as it cannot provide sufficient starlight suppression for imaging Earth-like exoplanets, the standard of which is considered to be $10^{-10}$ fainter than and 100 milli-arcseconds angular separation from its host star. At least, it cannot provide sufficient suppression with a reasonably sized starshade.

{\bf Morgan:} Were there mission ideas that predate the starshade?

{\bf Urania:} Yes Morgan, there were few such ideas. The first mention of an occulter with a space telescope cited is due to Lyman Spitzer in the 1960's.\cite{Spitzer_1962}.

In the 1980's, during the formulation of what is now called the Hubble Space Telescope, a number of very creative mission ideas for exoplanet detection were developed and bear some relevance to our story. One proposal was for an apodized telescope\cite{Banderman}. Another concept for a co-orbital occulting sphere (co-orbiting with Hubble) was also discussed\cite{Cocks}. A mission concept from the late 1990's, called UMBRAS, was an occulting body, but was not apodized\cite{UMBRAS_99}.

In the early 2000's there was an occulter concept, called BOSS, which was to work with the Next Generation Space Telescope (now called the James Webb Space Telescope) which used an apodization on a transmissive film \cite{BOSS}.

{\bf Morgan:} A binary structure sounds much easier to build and maintain than a transmissive film.

{\bf Urania:} Yes Morgan, it is significantly less complicated and better performing than a semi-transparent apodized optic. While all interesting ideas, none of these concepts were able to demonstrate a feasible design that could achieve the contrast needed for exoplanet detection.

{\bf Morgan:} So the real breakthrough was the apodization that allowed a very deep shadow?

{\bf Urania:} Yes Morgan, that was the real change: the first high-performance apodization solution that could be achieved with an opaque screen. This discovery kicked off the modern era of the starshade. Cash\cite{Cash_2006} provided an analytic solution, called the offset hypergaussian, that for the first time provided sufficient suppression with a starshade of practical size.

{\bf Morgan:} Offset hypergaussian? I haven't heard that term before.

{\bf Urania:} The offset hypergaussian is given by the apodization function
\begin{equation}
    A(\rho) = \begin{cases}
    1, & \rho < a\\
    e^{-\left(\frac{\rho-a}{b}\right)^n}, & \rho \ge a \,,\\
    \end{cases}
    \label{eq:hypergaussian}
\end{equation}
with tunable parameters $a,b,n$ (typically, $a=b=R/2$, $n=6$). The analytic solution lends itself well to examining the sensitivities and tolerances of the petal shapes, as done in Ref.~\citenum{Cash_2011}. The hypergaussian design also has an extremely wide wavelength coverage, as the performance improves with shorter wavelengths -- in fact suppression scales steeply as $N^{-2n}$, with $N = R^2/(\lambda Z)$ being the Fresnel number of a starshade of radius $R$ at a separation $Z$ from the telescope and at wavelength $\lambda$. The downside to this design is that it achieves its broadband response by having very narrow petal tips and valleys between petals. For a 30-m diameter starshade, the tips and valleys must be on the order of 10s of microns, which make the starshade hard to stow and deploy.

Vanderbei et al.\cite{Vanderbei_2007} developed a method, through numerical optimization, to achieve the same performance with a slightly smaller diameter starshade and with wider tips and valleys (on the order of 100s of microns). The wider tips come at the cost of a smaller bandpass, with performance degrading at shorter wavelengths, but the numerical optimization scheme allows one to suit the design to their needs and minimize the diameter and maximize tip widths while achieving the target performance over a desired bandpass.

{\bf Morgan:} How many petals are needed? What determines the length of the petals versus the diameter of the inner disk?

{\bf Urania:} The number of petals is set by how well they approximate the radial apodization function, generally $\gtrsim12$ petals are needed. Too few petals and the approximation begins to break down; higher order Bessel terms in the Jacobi-Anger expansion become relevant at larger radii\cite{Vanderbei_2003}, meaning the size of the deep shadow begins to shrink from the outside in. The ratio of the petal length to inner disk diameter is a design decision to be made for specific missions. Shorter petals are easier to make sturdy and require less stiffening structure per petal, but a larger number of petals are needed to approximate the apodization function.

{\bf Morgan:} You've mentioned the terms \emph{contrast} and \emph{suppression} a few times, what do these mean in the context of starshades?

\begin{figure}[!ht]
\begin{center}
    \includegraphics[width=\linewidth]{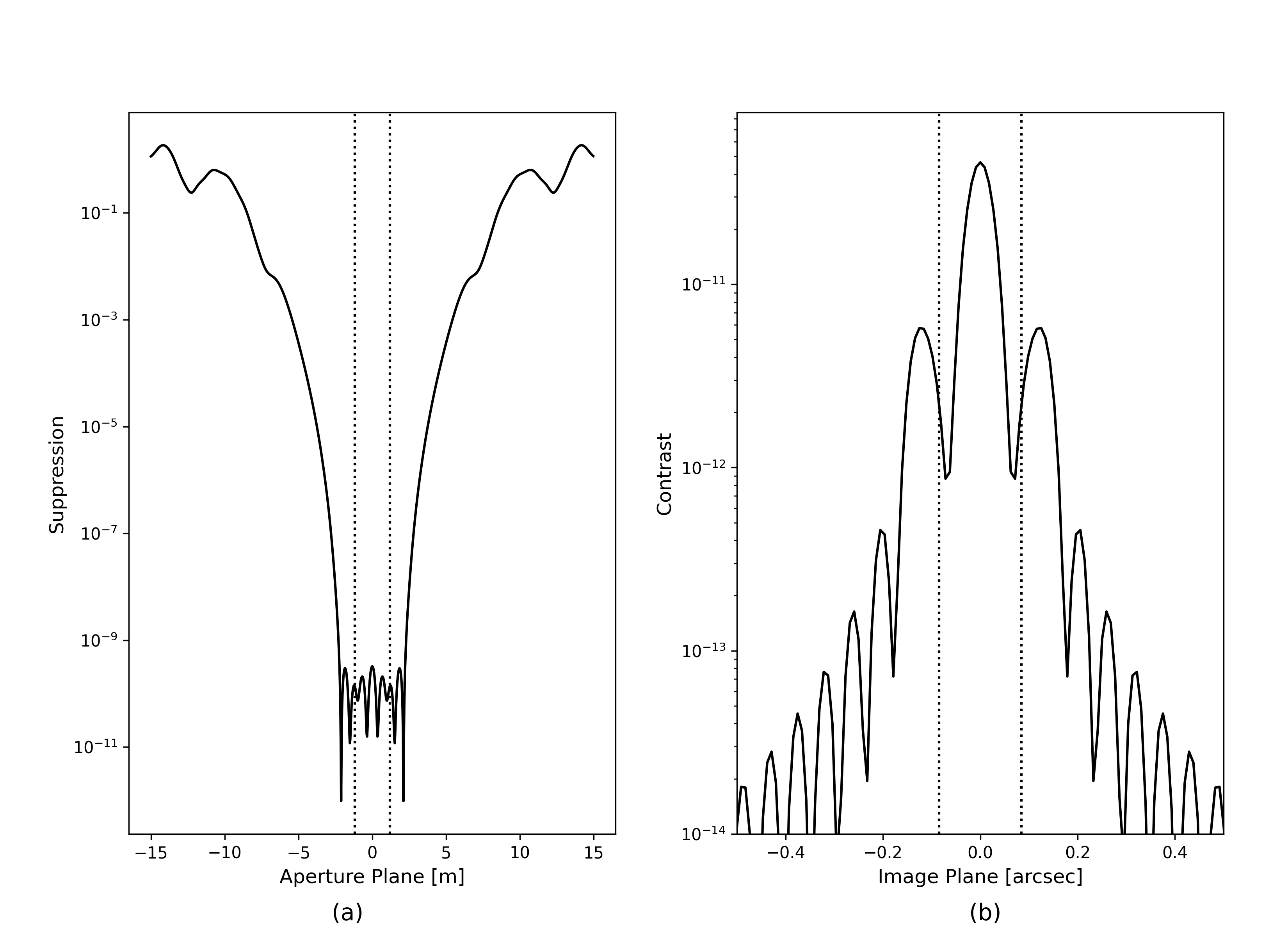}
	\caption{Suppression (a) and Contrast (b) for the Starshade Rendezvous Mission\cite{SRM}. Suppression is measured at the aperture plane. The vertical dotted lines in (a) mark the width of the telescope aperture. Contrast is measured in the image plane. The vertical dotted lines in (b) mark the inner working angle.}
	\label{fig:contrast_suppression}
\end{center}
\end{figure}

{\bf Urania:} Suppression is the total amount of starlight blocked by the starshade and is independent of the telescope's optics; it is usually quoted as the amount of residual light incident on the telescope's aperture. Contrast is measured in the image plane of the telescope and is thus dependent on the telescope and starshade properties. At a given spot in an image, the contrast is the amount of light there, relative to the on-axis peak brightness of the unocculted star. Contrast is generally the performance metric used in science contexts as it is tied to the ability to detect an object of a certain flux ratio at a given spot in the image. Figure~\ref{fig:contrast_suppression} shows a comparison between suppression and contrast measurements for the Starshade Rendezvous Mission\cite{SRM}.

{\bf Morgan:} I have another basic question: how big and how far in front of the telescope should the starshade be?

{\bf Urania:}  There are three main constraints driving the design of a starshade architecture\cite{Arenberg_Cash, Arenberg_2007, Glassman_2009}. To first order, you need a starshade to be large enough to create a deep shadow that covers the entire telescope aperture, with some additional margin. This additional shadow size (typically $\sim$2 m in diameter) gives sufficient tolerance to allow a formation flying solution to be found. In effect, the extra shadow size allows imperfect alignment between the starshade, its shadow, and the entrance aperture of the telescope. As we discussed earlier, the central disk of the starshade creates the geometric shadow to cover the telescope and the petals work to keep the diffracted light out of the center and create a deep shadow. The size of the deep shadow depends on many parameters of the starshade design and sets the first design constraint:
\begin{equation}
    R > \frac{D}{2}\frac{1}{\alpha(N, S, \Omega)}\,,
    \label{eq:1st_cons}
\end{equation}
where $R$ is the radius of the starshade; $D$ is the diameter of the telescope; $\alpha(N, S, \Omega)$ is the ratio of the radius of the deep shadow relative to the starshade radius, which depends to first order on the Fresnel number $N$, the target suppression level $S$, and the starshade petal shape $\Omega$. Typically, $\alpha$ is in the range of 0.05 - 0.15\cite{Glassman_2009}. So, say we want to build a starshade for a 2.4 m telescope and we think we can design a shape with $\alpha$ = 0.1, let's also add another 1 meter margin for formation flying, then we would need a starshade of radius $R\sim$ 13 m.

The main driver to the separation (Z) of the telescope and starshade is the inner working angle (IWA), which is the closest angular distance to the star we want to observe and is set by our science goals. Unlike coronagraphs, for the starshade the IWA is mostly decoupled from the size of the telescope and is set by geometry: IWA = $R/Z$. There is a small increase to the angular extent of the IWA due to the point spread function (PSF) of the telescope blending in scattered and diffracted light from the starshade's edge. This gives the IWA as
\begin{equation}
    IWA = \frac{R}{Z} + \beta\frac{\lambda}{D} \,,
    \label{eq:iwa}
\end{equation}
where $\beta$ is the fraction of the PSF width that we must move away from the starshade edge before we reach our target contrast\cite{Arenberg_2007_SPIE}. $\beta$ is set by the brightness of residual light at the edge. We can rewrite the second term of this equation in terms of the number of resolution elements across the starshade radius $n_{res} = (R/Z) / (\lambda/D)$; typically, $\beta <1$ and $n_{res}\sim 2$. This yields our second design constraint:
\begin{equation}
    Z > \frac{R}{IWA}\left(1 + \frac{\beta}{n_{res}}\right) \,.
    \label{eq:2nd_cons}
\end{equation}

As we move the starshade farther away from the telescope to decrease the IWA, the Fresnel number decreases and it becomes more difficult to control the diffraction, so we must counteract this by increasing the radius, which increases the IWA, and so on. This game is played until we reach a solution that satisfies all the constraints. The Fresnel number sets our third and final design constraint:
\begin{equation}
    \frac{R^2}{\lambda Z} > N_0(\Omega, S) \,.
    \label{eq:3rd_cons}
\end{equation}

$N_0$ is the minimum Fresnel number for which the starshade can sufficiently control the diffraction to provide the necessary suppression level $S$. $N_0$ is a shallow function of contrast (conversely, the contrast is a steep function of Fresnel number) and depends on the starshade petal shape. In practice, this number can be difficult to determine for numerically designed starshades, but typically $N_0\sim8-10$. For a hypergaussian design, it can be shown\cite{Cash_2011} that $N_0\sim 2S^{-1/12}$, so for a suppression of 10$^{-10}$, $N_0\sim13.5$.

\begin{figure}[!ht]
\begin{center}
    \includegraphics[width=\linewidth]{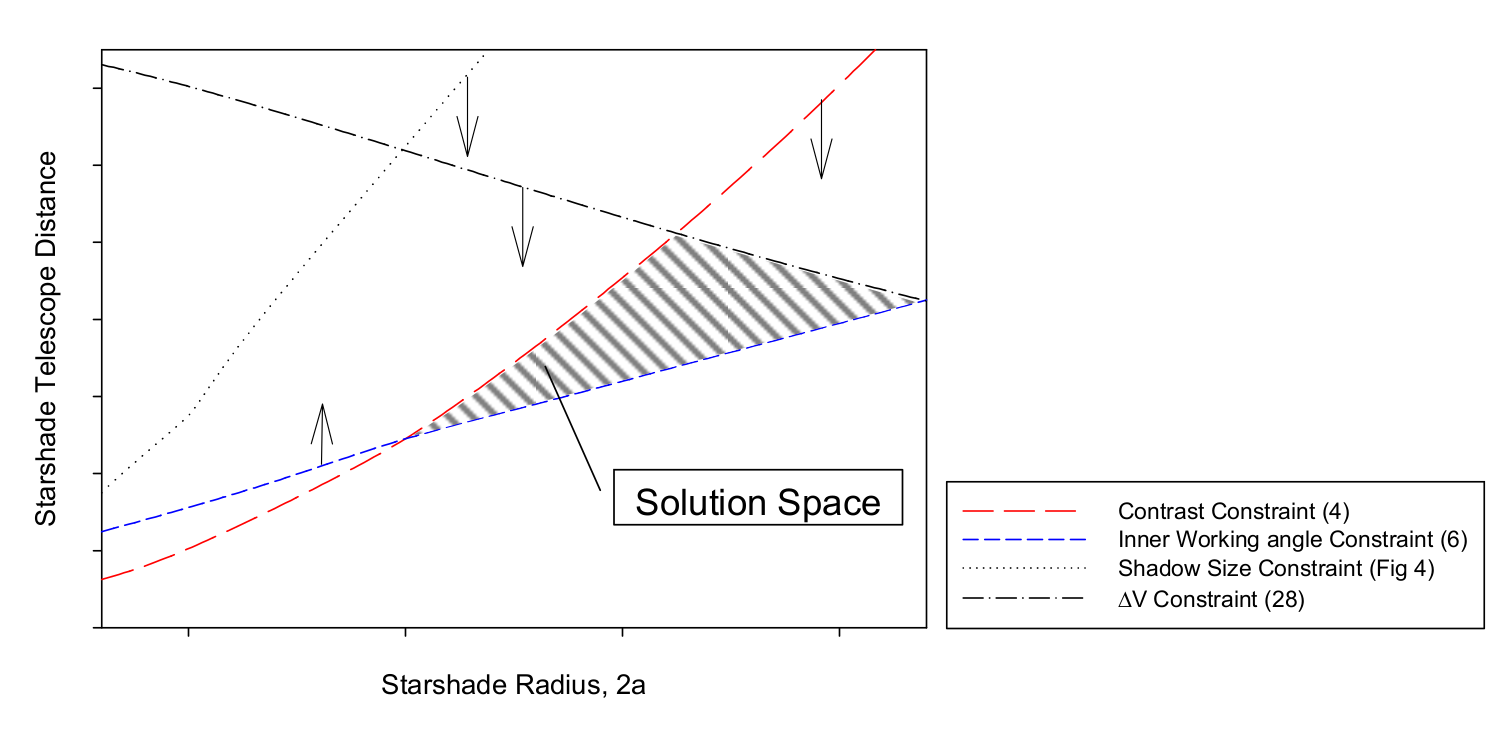}
	\caption{A typical design space plot for a Starshade, Figure~5 of Ref.~\citenum{Arenberg_2008_b} The equations referenced in the plot legend are the relevant references in the original paper.}
	\label{fig:solution_space}
\end{center}
\end{figure}

These various constraints can be laid out on a plane where starshade size is one axis and the Z distance is the other, the (R, Z) plane\cite{Arenberg_2008_b}. An example is shown as Figure~\ref{fig:solution_space}. In this plot, the various constraints are shown and the direction on the plane where they are satisfied are indicated by the arrows. The hatching shows the region of the (R, Z) plane where all of the constraints are satisfied.
%

{\bf Morgan:} So if I have a science goal that needs a certain contrast level at a given IWA, I can make a plot similar to Figure~\ref{fig:solution_space} and get a quick estimate of the starshade size and distance?

{\bf Urania:} Yes, you have it.  You seem to be getting the hang of this!

{\bf Morgan:} What is the ``shadow size constraint'' in Figure~\ref{fig:solution_space}? It doesn't seem to be a driving constraint.

{\bf Urania:} In a word, it isn't. For the case in Figure~\ref{fig:solution_space}, the shadow constraint of Equation~\ref{eq:1st_cons} is not the driving or limiting constraint. In systems design there are usually many factors that are related to a system parameter. The term driving requirement is given to the requirement that is the most stringent or sensitive. In most starshades for exoplanets Equation~\ref{eq:2nd_cons} and Equation~\ref{eq:3rd_cons} are what set the limit. Fortunately, there is always a solution to these constraints, at least mathematically. \cite{Arenberg_2007}

{\bf Morgan:} If telescope size isn't a main driver, how does the starshade size scale with the telescope size? Is it 1:1? In other words, to go from a 4 meter telescope to a 12 meter telescope, do I need a starshade that is 3$\times$ larger?

{\bf Urania:} Not exactly. It helps to think of the extra telescope size as an additive, rather than multiplicative, term, since most of the starshade size is used to generate the shadow in the first place. Once we have a starshade large enough to create a dark shadow, making the shadow larger requires (to first order) a linear addition to the starshade size. As we saw in Figure~\ref{fig:solution_space}, the main driver to the starshade size is IWA. If we have designed a 50 meter starshade to work at a given IWA for our 4 meter telescope, widening the shadow by 8 meters would result in a $15-25\%$ increase in the starshade size.

If we keep the IWA constant, we get additional help in keeping the starshade size reasonable when moving to a larger telescope. We can rewrite the Fresnel number as $N=R\theta/\lambda$, where $\theta$ is the IWA. If we hold $\theta$ constant, the Fresnel number increases with increasing starshade radius. A larger Fresnel number means it is easier to control the diffraction, so we can design an apodization function that can support a larger telescope relative to the starshade size.

{\bf Morgan:} Earlier, you mentioned that the IWA is decoupled from the telescope size and that telescope size is not a driving requirement. What role does the telescope size play in the system design?

{\bf Urania:} Of course the telescope size is driven by light collecting power (exoplanets are very faint!), but in terms of high contrast, high image resolution helps to separate diffracted starlight from the location of the exoplanet in the image. All of the diffracted light in the image emanates from the location of the starshade in the image (think of the Geometric Theory of Diffraction\cite{Keller_1962}), which means the contrast rapidly improves as you move away from the geometric IWA in the image. The amount to which we resolve the starshade can be expressed as the number of resolution elements across the starshade radius: $n_{res}=\frac{R/Z}{\lambda/D} = \frac{\theta D}{\lambda}$. An increase in resolving power allows us to relax the starshade performance as we can take advantage of the telescope's PSF to separate the diffracted starlight from the region of interest in the image.

As a result, we can allow tolerances on the starshade to be reduced by designing a starshade with a smaller geometric IWA (requiring a larger, farther starshade) and allowing the contrast to fall with the telescope's PSF. Or, if the IWA is held constant, the starshade is more and more resolved with increasing telescope size. If we are able to achieve the same contrast with poorer suppression, we can create a wider, shallower shadow in which to fit a larger telescope. This behavior enables {\bf Peretz et al.} to pair a starshade only 100 meters across with the 39 meter aperture of the next generation of extremely large telescopes.

{\bf Morgan:} Will there be an outer working angle?

{\bf Urania:} The starshade has no outer working angle (OWA). A coronagraph's OWA is usually set by the number and density of actuators on the deformable mirror. Because a starshade does not require wavefront control with a deformable mirror, its OWA can be unlimited, though the instrument's field of view may play a role. However, the field of view constraints can be overcome by offsetting the telescope's pointing; unlike a coronagraph, there is no requirement for the star to be on-axis.

{\bf Morgan:} Why doesn't the star need to be on-axis?

{\bf Urania:} Since the starshade creates the shadow independent of the telescope, it doesn't matter where the telescope is pointed as long as it remains in the shadow. As always, imagining yourself in an eclipse is helpful here. This freedom enables a wide field of view by building a mosaic image and also opens the door to multi-object spectroscopy.

{\bf Morgan:} If the starshade is located quite far from the telescope, won't it take forever to slew the starshade to a new target? How can I do a survey of many targets with a starshade?

{\bf Urania:} Now, now Morgan, forever is not a very precise term. But your intuition is good here. The starshade is of course far away, so moving between targets takes a while. Moving between targets requires thrust to change the velocity, $\Delta V$, of the starshade and consumes fuel. It is the finite fuel capacity that is a strong limitation of starshade missions.

{\bf Morgan:} Is there a solution to this problem?

{\bf Urania:} This issue was recognized very early in the development of the starshade idea. In this era, it was the fashion for missions to try and do it all themselves. So early mission analysis studied this mission planning\cite{Savransky_2010, Glassman_2011, Kolemen_2012, Soto_2019}.   Additionally, mission designs with multiple starshades were studied. \cite{Hunyadi}

This is of course an area of great concern. The ability to move the starshade to a target, make an observation and then return is key to establishing orbits. There are papers in this issue ({\bf Flinois et al.} and {\bf Morgan et al.}) that show the most recent thinking.

{\bf Morgan:} Are you saying that a Starshade can be be used as a survey tool?

{\bf Urania:} It can of course, but that is not its real strength. It is far better at characterization.

{\bf Morgan:} So it is a good idea to know about the targets ahead of time?

{\bf Urania:} Yes, that is a main point of the paper by {\bf Morgan et al.} in this issue.

{\bf Morgan:} Earlier, when we were talking about outer working angles, you mentioned wavefront control. Does a telescope working with a starshade need active wavefront control?

{\bf Urania:} No and this is the key strength of the starshade: the starlight has already been suppressed before it even reaches the telescope. Any distortion of the telescope's PSF due to wavefront errors introduced by its optics will occur at lower intensity levels in the wings of the PSF, which will have minimal impact on the exoplanet image -- the on-axis and off-axis PSFs are of equal strength. The quality of telescope's optics merely affect the image quality (Strehl ratio) as it would in standard imaging applications. By comparison, in a coronagraphic system, the full strength starlight enters the telescope and wavefront-induced distortions in the wings of the on-axis PSF can be many orders of magnitude brighter than the peak of the off-axis PSF.

{\bf Morgan:} Ok, I think I am starting to understand this. Does this mean starshades can work even outside a vacuum?

{\bf Urania:} Yes, excellent thought! Starshades work outside a vacuum, which makes it much easier to perform optical tests on the ground. The first starshade experiments\cite{Leviton_2007, Schindhelm_2007} were conducted in vacuum and air and saw little difference in performance between the two (except for light scattered by aerosols). This revelation, with subsequent explanation\cite{Cash_2011}, opened the door to experiments at larger separations in order to test at flight-like Fresnel numbers. Starshades have been tested over kilometers of separation on dry lakebeds\cite{Glassman_2013, Glassman_2014, Smith_2016} and with a heliostat on Kitt Peak\cite{Novicki_2016, Harness_2017}. Similarly, the optical experiments of {\bf Harness, et al.} were conducted in air.

{\bf Morgan:} That's all of the questions I have for today. I think I shall dream of hypergaussians tonight.

\subsection{Day Two: Starshade Technology and Engineering}
{\bf Morgan:} Yesterday was very helpful in getting me up to speed with how the starshade operates, but now I have some questions on the technologies needed to build and operate a starshade. First off, how do you get such a large structure in space?

{\bf Urania:}  Starshades are large structures that must fit into the fairly small volume of a fairing. Stowing and then deploying a starshade is a key element of its design and has been addressed with various designs over time.  \cite{Thomson_2010}

\begin{figure}[!ht]
\begin{center}
    \includegraphics[width=\linewidth]{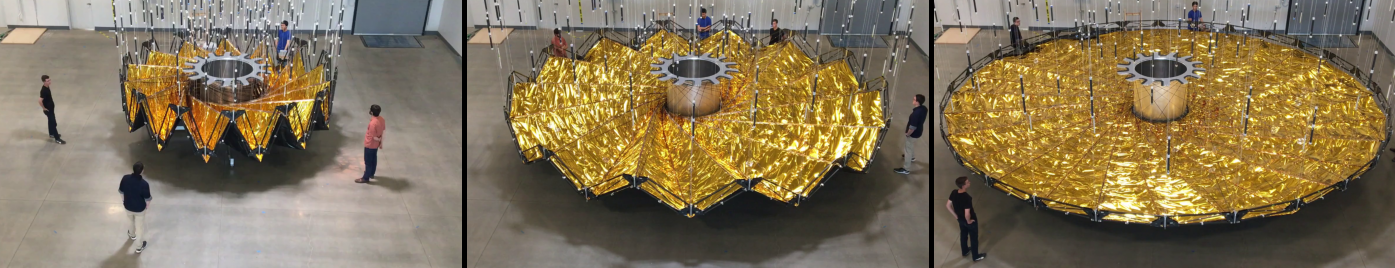}
	\caption{Deployment of 10 meter Inner Disk Subsystem prototype hardware. Photo credit: Jet Propulsion Laboratory.}
	\label{fig:deployment}
\end{center}
\end{figure}

{\bf Morgan:} How can we be sure that it will unfold properly?

{\bf Urania:} The development of dependable deployment mechanisms have been a major focus of starshade engineering\cite{Lillie_2008, Thomson_2010, Kasdin_2011, Webb_2014, Webb_2019}. The S5 technology program presented in {\bf Willems and Lisman} has a number of technology milestones\cite{Webb_2020, Arya_2020} that demonstrate the deployment of full-scale subsystems with sufficient accuracy and repeatability, and {\bf Arya et al.} present their solution to this design problem. Figure~\ref{fig:deployment} shows a sequence of such a deployment.

{\bf Morgan:} That looks hard...

{\bf Urania:} Of course, but there are many complex deployable systems that have been built and successfully operated over the last 60 years. Current starshade designs leverage much of this heritage.

{\bf Morgan:} Once a starshade is deployed, do we have to worry about vibrations?

{\bf Urania:} The starshade's optical performance is only sensitive to in-plane perturbations to its shape. Vibrations from thruster firings or low frequency vibrations from thermal effects that cause in-plane shape deformations are accounted for in the error budget\cite{Shaklan_2010}, however, they are expected to be minimal. Dynamic modeling shows that vibrations due to thruster firings dampen to micron-levels within a few seconds of firing\cite{Shaklan_2015}. Ref.~\citenum{Glassman_2016} performed a dynamical and thermal analysis on the optical performance and found the effects to be benign. Refs.~\citenum{Arya_2020, Webb_2020} have shown that the on-orbit thermal and mechanical stability requirements of critical components can be met.

{\bf Morgan:} What are the alignment tolerances between the starshade and telescope? Is it similar to an interferometer?

{\bf Urania:} No, it is much easier. Since the starshade suppresses starlight independently of the telescope (you can even achieve high contrast with the naked eye!), the alignment tolerances are measured in meters rather than wavelengths. The lateral position requirement is just to keep the telescope inside the deep shadow. The size of the deep shadow is designed to be larger than the telescope ($\sim2$ meters larger in diameter), which allows the starshade to freely drift around the line of sight to the star and need minimal corrections (every 100's of seconds) to keep the shadow centered over the telescope\cite{Flinois_2020}. The axial positioning requirement is even less stringent, as the shadow is created over many 100's kilometers axial distance. In the relatively benign gravitational environment of L2, it is understood to be straightforward to maintain $\pm1$ meter position control of the spacecraft and many methods to doing so have been identified\cite{Noecker_2007, Leitner_2007, Martin_2015, Scharf_2016, Flinois_2018_S5, Koenig_2019}. Much more stringent formation flying is achieved in the more onerous gravity environment of Low Earth Orbit with the International Space Station.

{\bf Morgan:} Yes, I am familiar with formation flying with the ISS, but that is done over a very short distance where you can easily see the target. How do you know where the starshade is over such large separations?

{\bf Urania:} Ah yes, you've highlighted the crux of the formation flying problem: accurately sensing the relative position over very large distances. The required $\sim 10$ cm positioning knowledge over 10,000's of kilometers separation corresponds to a sub-milliarcsecond angular measurement and would push on the resolution capabilities of most instruments. However, with the starshade we are able to exploit the light diffracted around the starshade to provide high precision position information over arbitrarily large distances.

{\bf Morgan:} But I thought we suppress all the diffracted light?

{\bf Urania:} We suppress the diffracted light at wavelengths in the bandpass of our design, but outside of that the starshade's performance quickly deteriorates and a strong signal of diffracted light emerges as the Spot of Arago. This diffraction pattern provides a one-to-one mapping of the lateral position of the starshade relative to the line of sight to the star\cite{Noecker_2007}. A pupil imaging sensor can image the diffraction pattern incident on the telescope's aperture and extract the lateral position of the starshade to high accuracy. This method of position sensing and formation flying control has been demonstrated to the necessary levels in multiple experiments\cite{Bottom_2020, Flinois_2020, Palacios_2020}.

{\bf Morgan:} If the Spot of Arago reappears outside the operating bandpass, does that mean light at that wavelength is constructively interfering? Like the circular disk we discussed yesterday?

{\bf Urania:} Exactly. It is easiest to think about this at long wavelengths -- the long wavelength light cannot discern the subtle shape of the petals, so the starshade begins to resemble a circular disk where the diffracted light reaches the telescope in phase.

{\bf Morgan:} How do we know the starshade will perform this way? Isn't the mathematics of diffraction complicated?

{\bf Urania:} There are a number of approximations we can employ to simplify the math into a manageable problem, for example, assuming that the starshade is large enough and far enough away that we can use scalar diffraction theory. We have developed a number of methods to solve the diffraction equation for the starshade\cite{Cash_2011, Cady_2012, Harness_2018, Harness_2020}, most recently in this issue with the work of {\bf Barnett}.

{\bf Morgan:} How confident are we that the math is correct?

\begin{figure}[!ht]
 \begin{center}
   \includegraphics[width=\linewidth]{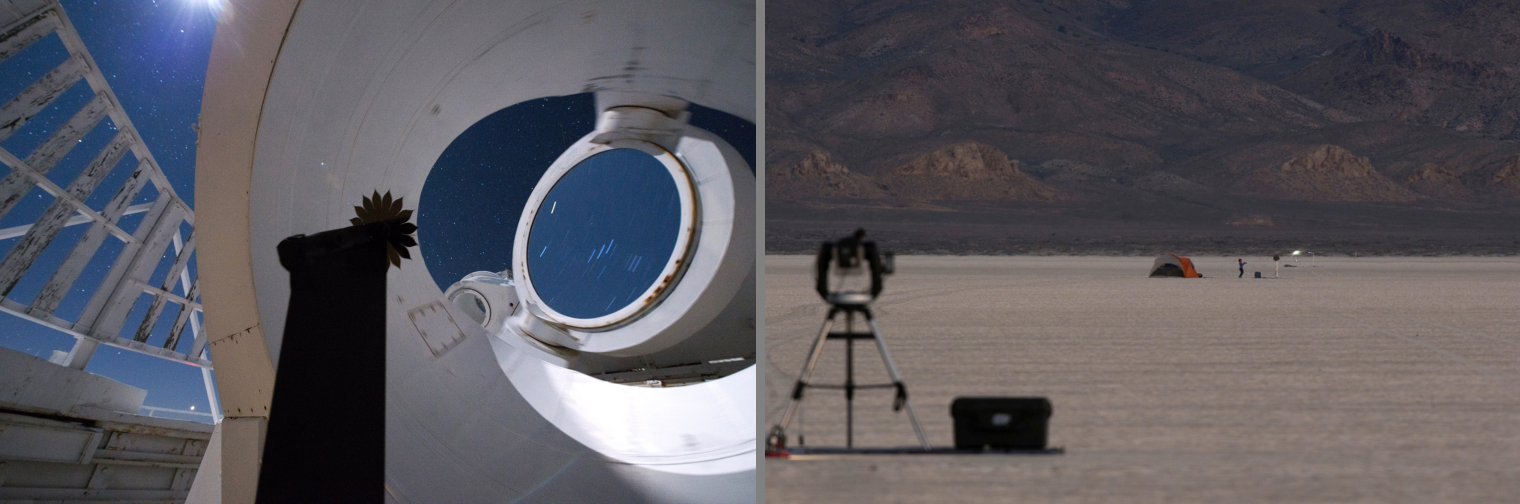}
	\caption{Left: A starshade is used to observe stars with the heliostat at the McMath-Pierce Solar Telescope\cite{Novicki_2016, Harness_2017}. Right: A starshade is tested over kilometers of separation on a dry lake bed\cite{Glassman_2013, Glassman_2014, Smith_2016}. Photo credit: Robert Brown (Northrop Grumman)}
	\label{fig:field_test}
\end{center}
\end{figure}

{\bf Urania:} There have been numerous efforts to test scaled down versions of starshades in order to validate the diffraction models. Although it is difficult to test at the large separations needed to achieve a flight-like Fresnel number, we have found a number of creative solutions, a couple are shown in Figure~\ref{fig:field_test}. These tests have been done in a lab setting over tens of meters\cite{Leviton_2007, Schindhelm_2007, Rocco_2009, Rocco_2010, Cady_2010, Sirbu_2013, Harness_2021}, on a dry lake bed over hundreds of meters\cite{Glassman_2013, Glassman_2014, Smith_2016}, and using a heliostat over kilometers\cite{Novicki_2016, Harness_2017}. The recent lab experiments of {\bf Harness et al.}, performed at a flight-like Fresnel number, show good agreement with the diffraction models.

{\bf Morgan:} How sharp do the petals' edges need to be?

{\bf Urania:} The large perimeter of the starshade petals provides a large source of scattered light\cite{Arenberg_2007_SPIE} from the brightest object in the sky, the Sun, which is 56 magnitudes brighter than a 30$^\text{th}$ magnitude exoplanet. With the target star suppressed, sunlight scattered off the petal edges becomes the dominant source of noise in the image\cite{Hu_2021_noise} and can have an integrated brightness of $V\sim25$ magnitude\cite{Shaklan_2021}. To minimize the scattering cross-section seen by the sun, the optical edges of the petals are designed to be as sharp as possible. The technical difficulty here is to create an optical edge that is as thin as possible, but is strong enough to maintain its shape and survive launch stresses.

{\bf Morgan:} If it is difficult to build a thin optical edge, why not reduce the reflected light by painting it black?

{\bf Urania:} The light scattered by sharp, uncoated edges is dominated by specular reflection and diffraction. A black coating can decrease the specular reflection component, but the coating increases the radius of curvature of the edge and leads to a stronger diffuse reflection. The magnitude of scattered sunlight from various edge geometries has been quantified through stray light models and laboratory experiments\cite{Arenberg_2007_SPIE, Casement_2012, Martin_2013, Casement_2016, Steeves_2016, Steeves_2018, Hilgemann_2019, Hilgemann_2019_S5, McKeithen_2019}. In this issue, {\bf McKeithen, et al.} report on the promising development of anti-reflection coatings that are thin and able to reduce the scattering brightness by at least an order of magnitude.

{\bf Morgan:} Do we have to worry about other sources of light (e.~g.~moonlight) reflecting off of the telescope-facing part of the starshade?

{\bf Urania:} Light from secondary sources such as moonshine, Earthshine, the Milky Way, etc, can be reflected into the telescope\cite{Arenberg_2007_SPIE}. For this reason, the opaque optical shield on the telescope side is made from a diffuse scattering, low reflectivity material such as Kapton. Ref.~\citenum{Shaklan_2015} estimates the worst case incident angle for Jupiter reflecting off Kapton makes the starshade appear as 30$^\text{th}$ magnitude in integrated brightness. However, the power is spread over the size of the starshade and most of it comes from the central disk of the starshade, which is a number of resolution elements away from the IWA. The effect of secondary stray light sources are estimated in {\bf Hu, R. et al. -- Noise Budget}.

{\bf Morgan:} Do we have to worry that the plume from the starshade's fuel will reflect light (e.~g.~sunlight) and contribute to the background noise on the science detector?

{\bf Urania:} Yes, the exhaust plume from the starshade's chemical thrusters creates a source of scatterers that can reflect sunlight into the telescope and ruin the image. To prevent this, the telescope and starshade are coordinated to pause the observation while the starshade is thrusting to maintain alignment, which occurs in $\sim$500 second intervals. Ref.~\citenum{HabEx_Report} estimates the plume will exit a 10 arcsecond diameter field of view in just 0.8 seconds. Ref.~\citenum{Flinois_2020} developed a formation flying framework that aims to minimize the observing time lost to closing the shutter by minimizing the number of thruster firings and syncing the firings to occur at times when the camera electronics are already in the process of being read out.

{\bf Morgan:} What happens if a micrometeoroid hits the starshade?

{\bf Urania:} The starshade can tolerate a significant number of micrometeoroid-induced pinholes in its opaque optical shield before it begins to degrade in contrast. A cumulative area of 1 cm$^2$ of clear-through pinholes would result in only 10$^{-12}$ contrast\cite{Shaklan_2015}. Adding multiple layers to the optical shield will prevent most micrometeoroid impacts from creating a pinhole with a direct line of sight to the telescope\cite{Arenberg_2007_SPIE}, further reducing the effect of pinholes. Similarly, sunlight entering the opaque shield through the pinholes will require a significant number of bounces to reach the telescope and will be substantially reduced. Ref.~\citenum{Shaklan_2015} conservatively estimates the starshade can sustain 10's to 100's of square centimeters of pinholes while meeting its contrast requirements.

{\bf Morgan:} I've heard about proposals for spinning starshades -- why would the starshade spin during observations?

{\bf Urania:} The starshade is not required to spin during observations, but that is the design choice made for the potential Starshade Rendezvous Mission\cite{SRM}, where the starshade spins at $\sim0.33$ rpm. Spinning the starshade accomplishes three goals. First, spinning the starshade stabilizes its attitude and eliminates the need for reaction wheels for attitude control. Second, spinning the starshade keeps it rotating in and out of direct starlight and prevents thermal gradients from developing. The starshade is designed to be thermally stable and distortions to its shape from thermal gradients are expected to be low\cite{Glassman_2016, Webb_2020}, but spinning helps ease the thermal requirements of materials. Finally, spinning the starshade at a rate faster than the exposure time (which is the current baseline) will smear any defects on the starshade into a uniform annulus interior to the inner working angle in the image. A smooth, uniform annulus can more easily be subtracted from the image in post-processing and will not be mistaken for an exoplanet. If the starshade is spun at a rate slower than the exposure time, defects on the starshade will not be smeared into an annulus, but will still move in subsequent images while the astrophysical scene will remain stationary. This opens the door for an angular differential imaging (ADI) like post-processing method to distinguish defect from exoplanet.

{\bf Morgan:} All of this engineering talk is making my head spin, so let's break for the day.

{\bf Urania:} Agreed. Tomorrow we shall focus on observations and discuss the exciting science the starshade can enable.

\subsection{Day Three: Starshade Science}

{\bf Urania:} Good morning! So far we've learned the basics of how a starshade operates, along with the various technologies and engineering solutions that enable it. Today, let's examine starshade observations and their scientific output.

{\bf Morgan:} I'm ready!

{\bf Urania:} To start, what science questions would you like to answer?

{\bf Morgan:} I would like to find Earth's twin. A rocky, terrestrial exoplanet in the habitable zone of its host star.

{\bf Urania:} A pale blue dot, say?

{\bf Morgan:} Yes, and search it for biosignatures -- look for signs of life.

{\bf Urania:} So you would like spectra of that terrestrial exoplanet?

{\bf Morgan:} Yes! I'm searching for a pale \emph{rainbow} dot, if you will.

{\bf Urania:} Well, you've come to the right place. Starshades excel at spectral characterization of faint exoplanets.

{\bf Morgan:} What makes a starshade better than a coronagraph?

{\bf Urania:} To be fair, starshades and coronagraphs each have different strengths and weaknesses. For example, by minimizing the amount of starlight reaching the observatory, starshades are largely insensitive to optical aberrations within the observatory. Coronagraphs, however, require exquisite wavefront control to correct for the  aberrations that give rise to speckles when the starlight encounters imperfections in the observatory's optical system. A coronagraph's performance is also impeded by any telescope obscurations, such as the shadow of the secondary mirror and its support structure in the case of an on-axis telescope, while a starshade's performance is unaffected by such obscurations.

{\bf Morgan:} On Day One, you mentioned that the telescope's optics affect the image quality; I think a large obscuration would affect the telescope's PSF and reduce the image quality.

{\bf Urania:} Yes, that's correct. The obscurations do not affect the starshade performance, but they do affect the quality of the image collected by the telescope, although much less so than with a coronagraph.

{\bf Morgan:} OK, thanks for clarifying. Please continue.

{\bf Urania:} Without the need for a host of additional optics to control the wavefront, the optical throughput of the starshade observatory can be much higher than that with a coronagraph. In general, starshades' insensitivity to the details of the observatory mean that they allow for the detection of fainter exoplanets than coronagraphs for a given observatory design. Starshades offer another advantage in terms of exoplanet detection: they can support an IWA smaller than $1 \lambda /$D and they have no OWA. Coronagraphs, however, typically have a minimum IWA of $1 \lambda /$D, and an OWA that is determined by the number of actuators spanning the wavefront control system's deformable mirror. Starshades therefore support exoplanet observations over a greater area surrounding the host star.

Once a planet has been detected, starshades offer additional advantages compared with coronagraphs: they operate over a wide bandpass. Assuming a hypergaussian starshade design (see Ref.~\citenum{Cash_2011}), the starlight suppression $S$ is related to the wavelength $\lambda$, the starshade-telescope separation $Z$, the radius of the inner disk component of the starshade $a$, and two hypergaussian components ($b$ and $n$) as:
\begin{equation}
    S = \left [ n! \left ( \frac{\lambda Z}{ab} \right )^n \right ]^2
    \label{eq:hg_suppression}
\end{equation}
where this equation has been adapted from Ref.~\citenum{Arenberg_2007}. It can be recognized from this equation that the contrast is strongly dependent on the Fresnel number ($N$): $S\propto N^{-2n}$; typically $n=6$. Hence, the starlight suppression improves for shorter wavelengths where all other parameters are fixed. Hence, the starshade's size and separation from the telescope are chosen on the basis of the longest desired wavelength, and, in this ideal hypergaussian case, there is no lower limit to the starshade's bandpass.

{\bf Morgan:} These are all wins for the starshade so far...

{\bf Urania:} There are several programmatic considerations that tend to disfavor the starshade. First, as we discussed on Day One, a starshade takes time and fuel to slew between targets, reducing the total amount of time available for exoplanet observations. Because a coronagraph is inside the telescope, it does not require any additional slew time, so it can support a survey with a larger number of target stars than a starshade system for a given total survey time. Second, coronagraphs are a mature technology for ground-based exoplanet imaging, and will soon be demonstrated in space as part of the Coronagraphic Instrument (CGI) on the Nancy Grace Roman Space Telescope. Starshades, on the other hand, have not yet been demonstrated in a scientific context, due to the challenges associated with using starshades on the ground. Hence, there are some technical considerations, such as the starshade's deployment after launch, that will be fully realized only in space. Other considerations, such as maintaining the alignment between the starshade and telescope, can be at least partially tested on the ground ahead of launch, as we discussed yesterday.

{\bf Morgan:} So the coronagraph is fast, but has lower throughput and the starshade has higher throughput, but is slow -- it sounds like these instruments are complementary to each other.

{\bf Urania:} I agree. Coronagraphs are quick and nimble and can survey many targets in a short amount of time, but cannot support the spectral characterization as well as the starshade can. Starshades are slow to move between targets, but have excellent throughput and wavelength coverage. I think a coronagraph for surveying plus a starshade for characterizing makes for a powerful observatory.

{\bf Morgan:} If I use a starshade, what sources of noise will ultimately limit my ability to detect faint exoplanets outside the inner working angle?

{\bf Urania:} The sources of noise that will impact observations of faint exoplanets are summarized nicely in {\bf Hu, R. et al. -- Noise Budget}. The largest impediment to detecting Earth-like exoplanets is likely to come from exozodiacal dust around the target star, but for now I will focus on noise sources that are specific to a starshade mission (as such, I will also neglect local zodiacal dust and detector noise).

As we discussed yesterday, the largest source of noise from the starshade comes from solar glint off the petal edges\cite{Arenberg_2007_SPIE, Shaklan_2015}; In this issue, {\bf Shaklan et al.} discuss its expected effects and {\bf McKeithen et al.} offer a possible mitigation strategy. The solar glint appears as two bright lobes interior to the IWA and should remain constant for the duration of the observation. The ongoing Starshade Data Challenge summarized in {\bf Hu, R. et al. -- Data Challenge} should provide insight into our ability to calibrate and subtract these lobes from the image.

The next dominant source of noise comes from perturbations in the starshade's shape that degrade the suppression performance. Perturbations can occur during manufacturing and assembly, in-space deployment, or due to distortions caused by thermal gradients. The dominant types of perturbations have been identified and are estimated in Refs.~\citenum{Shipley_2007, Arenberg_2008, Shaklan_2010, Shaklan_2011, Cash_2011, Shaklan_2017}. In this issue, {\bf Hu, R. et al. -- Noise Budget} and {\bf Peretz et al. -- Performance Envelopes} estimate the effect of these perturbations on exoplanet detection.

{\bf Morgan:} I know that coronagraphic images suffer from residual ``speckles'', which result from uncorrected wavefront errors. Are these a problem for starshades as well?

{\bf Urania:} No. If you remember back in Day One, we learned how the starshade is not sensitive to wavefront errors in the telescope. Any stray light that enters the telescope has already been reduced by many orders of magnitude ($> 9$) and speckles that arise from wavefront errors will be even further reduced in the wings of the PSF. For coronagraphs, the speckles are generated from the full strength starlight and can be as bright or brighter than the planet light.

\begin{figure}[!ht]
\begin{center}
    \includegraphics[width=0.9\linewidth]{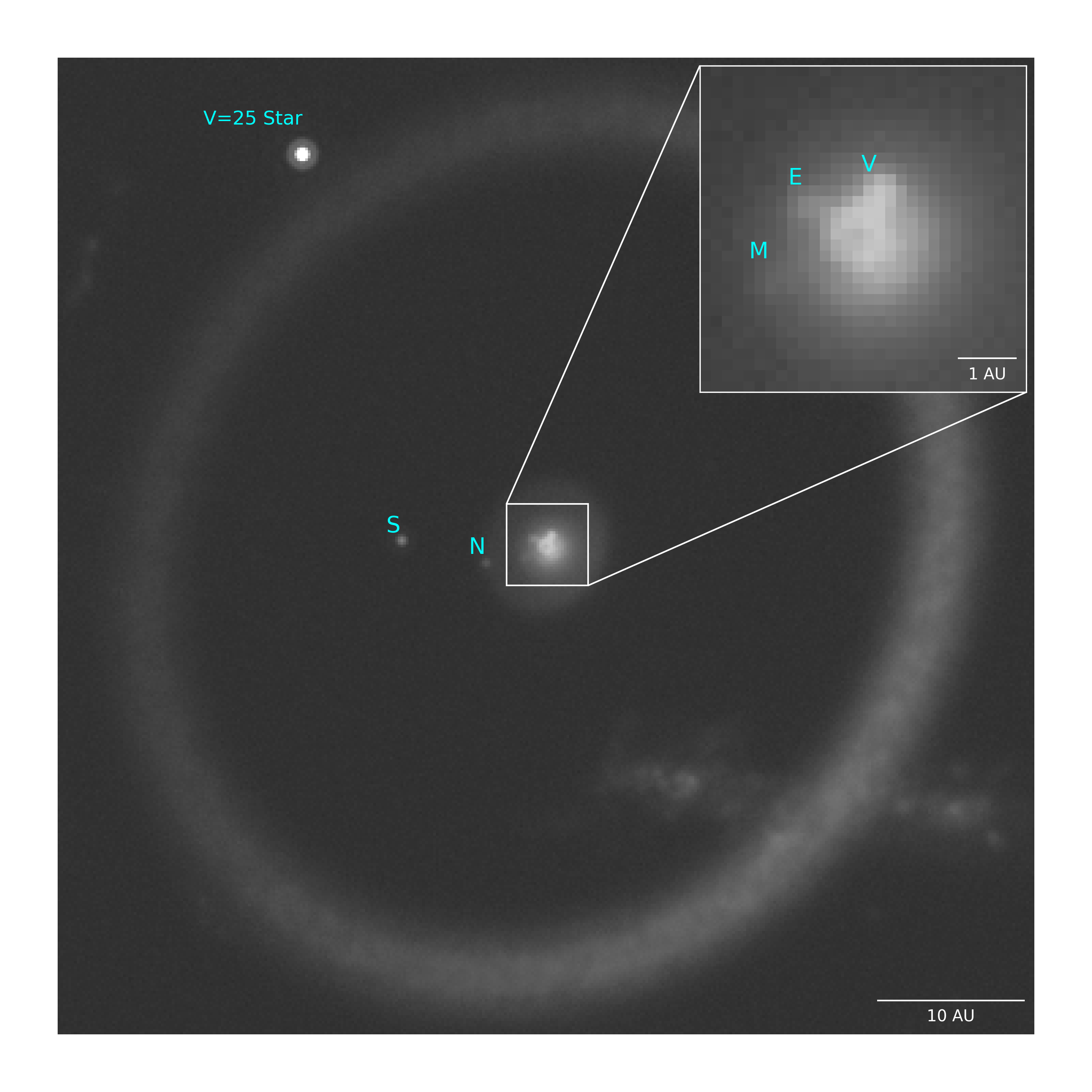}
	\caption{SISTER (see {\bf Hildebrandt et al.}) generated image of the star $\beta$ CVn in the Roman telescope `Green band' (615-800 nm). This fictional planetary system includes Venus, Earth, a `super Mars' (the diameter of Earth but at the location of Mars), a Neptune-sized planet, Saturn, 3$\times$ the solar system level of zodiacal dust, and a Kuiper belt analog debris disk. The system has a visual inclination of $30^{\circ}$, is dynamically stable, and is consistent with recent radial velocity analysis\cite{Kane_2020}. The instrumental effects include an imperfect starshade with 10$^{-10}$ contrast at the IWA, formation flying misalignments, solar glint, and optical throughput\cite{Romero-Wolf_2021} and detector noise\cite{Roman_detector} consistent with the Roman telescope end of life. The integration time is less than a week. A background galaxy can be seen in the lower right corner, along with a V=25 background star in the upper left. Simulation credit: Sergi Hildebrandt (JPL/Caltech).}
	\label{fig:sister}
\end{center}
\end{figure}

{\bf Morgan:} If we have not yet flown a starshade, how do we know what the images will look like?

{\bf Urania:} That's a great point. While it is true that we have not observed an astronomical object with a full scale starshade, the sub-scale experiments done by {\bf Harness, et al.} have validated diffraction models such that we are confident in estimating the optical performance. These diffraction models are incorporated into image simulations that allow us to explore what a real starshade observation would look like. {\bf Hildebrandt et al.} present the Starshade Imaging Simulation Toolkit for Exoplanet Reconnaissance (SISTER), which incorporates instrumental, astrophysical, and starshade noise sources into the simulation software.

Figure~\ref{fig:sister} shows a SISTER-simulated broadband (615-800 nm) image of a fictional planetary system around the star $\beta$ CVn (aka Chara). With an integration time of less than a week, this image shows the power of the starshade, as faint Earth-like planets are detectable even in the presence of astrophysical and instrumental noise. If you would like to better understand what a starshade image will look like and what the main limitations will be, you can download SISTER from \url{http://sister.caltech.edu} and try it for yourself!

{\bf Morgan:} Great! I will download that code immediately and start simulating starshade images on my own. I will want to take spectra of exoplanets over a wide range of wavelengths. What limits do starshades place on the spectral coverage?

{\bf Urania:} The starshade's wavelength coverage can be made arbitrarily large, but at the expense of a larger starshade at a larger separation. In general, a wider bandpass, or a redder bandpass means a larger, farther away starshade. On Day One, we discussed how the spectral coverage of the starshade is set by the design of its apodization function. Remember that the hypergaussian design\cite{Cash_2006, Cash_2011} has a wide bandpass, whose suppression scales with hypergaussian parameter $n$ as $\lambda^{-2n}$ ($n\sim6$). The size of the starshade is set by the reddest wavelength and the suppression improves for all wavelengths blue of that. This design however, results in narrow petal tips and valleys that are more difficult to stow and deploy. A numerically determined apodization function\cite{Vanderbei_2007} can selectively choose the bandpass to operate over and will result in a slightly smaller starshade with wider tips and valleys. The width of the tips and valleys are set by the size of the bandpass, with a wider bandpass requiring narrower tips and valleys.

{\bf Morgan:} Are there restrictions on the angular distance between the starshade and the Sun during science observations?

{\bf Urania:} Yes, to prevent sunlight from scattering off the starshade and into the telescope, the observatory has more stringent solar keepout zones than a typical space observatory. The angle between the starshade normal axis and the sun is kept between $40^\circ<\phi<83^\circ$ during science observations (see {\bf Shaklan et al. in this issue}). The lower bound is the standard observatory keepout zone set by the telescope's baffle, while the upper bound is set by the starshade $-$ at angles $>83^\circ$, the telescope-facing side of the starshade would be illuminated and scatter sunlight into the telescope.

{\bf Morgan:} What if the star is resolved?

{\bf Urania:} The angular size of the star has minimal impact on the starshade's performance; in fact, the starshade even works on resolved objects such as Jupiter\cite{Novicki_2016}. The main impact of a resolved source is to slightly reduce the size of the shadow, but this can be compensated for by designing a slightly larger starshade or tightening the formation flying performance. For a target with angular extent $\theta$, observed with a starshade at separation $Z$, the diameter of the shadow is reduced by $\theta Z$. For the HabEx starshade at 76,600 km, a 1 milli-arcsecond star reduces the size of the shadow by 0.37 m, which represents only 0.7\% of the starshade's 52 m diameter.

{\bf Morgan:} Starshades sound pretty great, but they also sound expensive and technically challenging! What's an example of a science case that's only possible to address with a starshade.

{\bf Urania:} The advantages of starshades compared with coronagraphs that we discussed earlier are such that a telescope would likely achieve a deeper contrast over a larger field of view and larger bandpass with a starshade than the same telescope would with a coronagraph alone. The Roman Space Telescope illustrates this concept: because the Roman telescope itself was donated to NASA by the National Reconnaissance Office, its secondary obscuration is less favorable than would have been the case if the telescope was originally designed with exoplanet imaging in mind. Hence, a ``starshade rendezvous'' concept, in which a starshade is launched and aligned with the Roman spacecraft after the launch of the telescope itself, has been proposed\cite{SRM}. The starshade would improve Roman's inner working angle by a factor of two or more, and improve the limiting contrast by an order of magnitude. While Roman's coronagraph would enable direct images of well-separated Jupiter-mass planets, the small IWA of the starshade would enable the direct imaging of a large diversity of small planets (rocky planets, mini-Neptunes, etc.) in the habitable zones of the nearest stars. The simultaneous large OWA also means that  well-separated planets down to Neptune's mass will be accessible for characterization. For larger telescopes that may be launched in the coming decades (such as HabEx or LUVOIR), the starshade increases the number of rocky, habitable-zone planets that smaller ($<8$-m) telescopes can detect and characterize compared with a coronagraph\cite{Stark_2016}.

{\bf Morgan:} What about other types of science? Are starshades only useful for exoplanet science?

{\bf Urania:} The starshade is useful for any high-contrast science application, not only exoplanets. For example, the starshade could be used to study the material surrounding active galactic nuclei, the dust shells of AGB stars, the Sun's corona, and even the faint rings of Jupiter!

{\bf Morgan:} What about the outflows of Wolf-Rayet stars?

{\bf Urania:} Yes! Very good suggestion! I will have to write that one down.

{\bf Morgan:} Here's another -- conduct a deep galaxy survey in regions of the sky normally blocked by bright stars.

{\bf Urania:} Great! Please feel free to keep thinking about different science applications for the starshade. We are always looking to expand the membership of the starshade community.

{\bf Morgan:} What kind of telescopes can work with a starshade? Must they be of a certain size or a certain quality?

\begin{figure}[!ht]
\begin{center}
    \includegraphics[width=\linewidth]{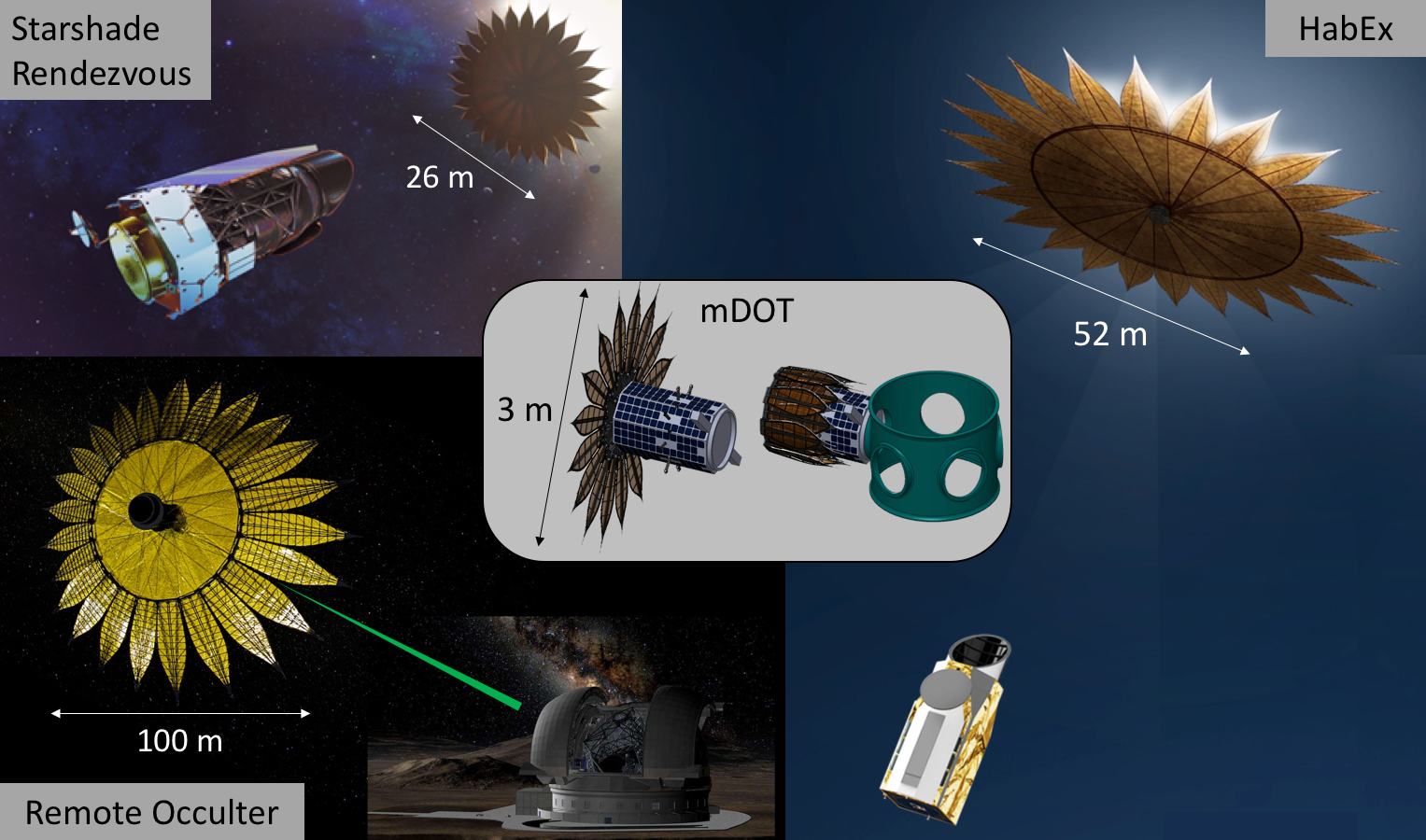}
	\caption{Various mission concepts, with a large range in telescope size, that propose using a starshade. mDOT\cite{DAmico_2019} uses a 10 cm telescope; Starshade Rendezvous\cite{SRM} operates with the 2.4 m Roman Telescope; HabEx\cite{HabEx_Report} proposes telescopes from 2.4 m to 4 m; and Remote Occulter\cite{Remote_Occulter} operates with a 39 m ground-based extremely large telescope. mDOT image credit: D'Amico, Macintosh, Koening, and NASA Ames. Other images were adapted from their respective citations.
	}
	\label{fig:arch}
\end{center}
\end{figure}

{\bf Urania:} Since the starshade creates a deep shadow independently of the telescope, almost any telescope size or design will support high contrast observations behind a starshade. However, in order to achieve high contrast at a certain angular separation, the size and quality of the telescope's PSF must be taken into consideration. Figure~\ref{fig:arch} shows the wide range of missions that are considering a starshade. The telescopes being considered for these missions range from the very small $-$ a 10 cm telescope on a small satellite\cite{DAmico_2019}, to the extremely large $-$ an orbiting starshade working with a 40-meter ground-based telescope\cite{Peretz_2019}. A small telescope paired with a small, close-in starshade could provide interesting disk science\cite{DAmico_2019} with a fast, nimble, and relatively inexpensive starshade. Larger telescopes require large starshades, but the improved angular resolution more easily separates diffracted light from the exoplanet of interest and allows us to relax the tolerances and performance requirements of the starshade.

There is one other attribute that is needed from the telescope, namely some sort of communication with the starshade.  Since this is not an optical property, one could say that it's a property of the observatory itself, whether that observatory is located on the ground or in space.

{\bf Morgan:} To close the alignment loop?

{\bf Urania:} That's correct, you are really getting this stuff! Just to be clear, I should mention that this kind of link is very hard to retrofit into a flying space telescope.

{\bf Morgan:} So the space telescope would have to be designed to work with a starshade.

{\bf Urania:} It would have to be designed into a space telescope, like Roman. But it could be a retro-fit for a ground-based observatory.

{\bf Morgan:} It sounds like the starshade is quite versatile.

{\bf Urania:} Indeed! We have only begun to scratch the surface of different starshade configurations. We could reduce the size and separation by observing at very blue wavelengths, or working at a moderate IWA; put a starshade on a heliostat to observe from the ground\cite{Novicki_2016, Harness_2017}; fly multiple starshades to cut down on slew time\cite{Hunyadi}; assemble starshades with...

{\bf Morgan:} OK! OK! Maybe that's enough for today.

\subsection{Day Four: Starshade Programmatics}

{\bf Morgan:} In the past three days, you've made a true believer of me and I'm now convinced I need a starshade.

{\bf Urania:} Excellent! Welcome to the club.

{\bf Morgan:} However, I'm still unsure of how we turn this into a reality. How can we validate the starshade's performance before launch? It sounds like it would be difficult to do from the ground, given gravity and the distance between the telescope and the starshade.

{\bf Urania:} This is a very good question. This is the thinking behind the starshade technology program (S5): to identify the necessary technologies and show them to be well understood and of sufficient determinism to allow a starshade mission. This is what is driving the technology development efforts presented in {\bf Willems and Lisman}.

{\bf Morgan:} What is the best example of this?

{\bf Urania:} In my mind the best example is the optical performance of the starshade. Given the size and distances needed for a flight scale starshade, the Earth is not large enough to do a full scale test. Therefore, we must use scaled testing. In order for this to be validated and be convincing, analytic work and experiments to validate the analysis are needed. Analyses of the starshade's shadow have been at the root of the planet-finding starshade from inception\cite{Arenberg_Cash, Cash_2006} to the present\cite{Harness_2021}.

{\bf Morgan:} Can we really expect to successfully launch a starshade to use with one of these future observatories without an initial space-based starshade technology demonstration?

{\bf Urania:} The question of what is required to show technical readiness for a starshade mission was addressed by a group assembled by NASA HQ in late 2015. This group was called the ``StarShade Readiness Working Group'' (SSWG) and it had a formal charter\cite{SSWG_charter}. The deliverable item this group was to produce was to ``Develop and deliver to the NASA Astrophysics Director by July 2016 a recommendation for a plan to validate starshade technology (to TRL~6/7) that is both necessary and sufficient prior to building and flying a Starshade Rendezvous science mission.''  This group was made up of representatives from various NASA Centers, JPL, academia and industry and considered many options to answer the questions:
\begin{enumerate}
    \item How do we go from Technology Readiness Level (TRL)~5 to TRL~6?
    \item Imagine ourselves at Key Decision Point (KDP)-C for a possible starshade science mission. Looking back, how did we convince all stakeholders to approve the mission?
    \item Put another way: Is a flight tech demo required to prove TRL~6, and if so, what is it?
\end{enumerate}

Here is what the SSWG found: first, a ground-only development strategy exists to enable a starshade science flight mission such as WFIRST (now Nancy Grace Roman Space Telescope) Starshade Rendezvous.  A prior flight technology demonstration is not required prior to KDP-C of WFIRST Rendezvous.  Development solutions exist that support a WFIRST Starshade Rendezvous by Launch Readiness Date fiscal year 2026-28. Additionally, the technology development for a Starshade Rendezvous mission is likely to provide significant technology benefits to both the HabEx and LUVOIR large mission studies The SSWG found that there were two optional enhancements to the recommended development approach. The first enhancement, a flight technology demonstration (mDOT\cite{DAmico_2019}), would enhance the ground development strategy for formation flying sensing and control and optical performance with additional cost and technical risk. The second enhancement was that long baseline ground demonstrations in air may provide some additional benefit for optical verification but at medium-to-high risk for interpretation of results\cite{SSWG_briefing}.
Simply put, the SSWG convinced itself and its built-in review committee that a scaled ground-only set of demonstrations are necessary and sufficient to achieve maturity for a flight mission.

{\bf Morgan:} I again have to say that starshades sound expensive and technically challenging.

{\bf Urania:} Yes, any space-based observatory will be expensive and technically challenging, and a starshade is not necessarily unique in those aspects. Current coronagraph technologies are driving the flagship missions towards large, expensive, ultra-stable telescopes\cite{HabEx_Report, LUVOIR_Report}. Since the starshade does not put strict requirements on the telescope, the cost and complexity shifts from the telescope to the starshade. Refs.~\citenum{Redding_2019, HabEx_Report} outline a starshade-only architecture for HabEx that forgoes the coronagraph and as a result has a lighter, more compact telescope. The science yield is lower than the hybrid architecture, but higher than that of the coronagraph-only architecture. Starshades are too new a technology and have not seen enough dedicated effort or funding to tell if they will be ``harder'' or ``more expensive''.

{\bf Morgan:} What are the prospects for using a starshade in conjunction with one of the upcoming telescopes (e.~g.~JWST or Roman)? How about observatories farther in the future like HabEx or LUVOIR?

{\bf Urania:} We touched on this briefly in Day Three, but there's a lot more to say! Ref.~\citenum{Soummer_2009} studied the possibility of a starshade rendezvous with JWST, but that ship has likely sailed. In the near term, the miniaturized Distributed Occulter Telescope (mDOT) proposes a small 3 meter starshade operating in low Earth orbit with a small-sat telescope that could provide interesting disk science and serve as a technology demonstration. In 2019, NASA released a report describing a starshade that would work in tandem with the Roman Space Telescope\cite{SRM}. The goals of this rendezvous would include technology demonstrations and science -- for example, searching for planets around the nearest 10-12 sunlike stars. Both the HabEx\cite{HabEx_Report} and LUVOIR\cite{LUVOIR_Report} concept studies considered a starshade component to their mission, but their final architectures are yet to be defined. HabEx has baselined a starshade in their primary mission architecture to provide spectral characterization of rocky planets in the habitable zone. Appendix I.1 of the LUVOIR final report discusses the possibility of a starshade rendezvous, but they conclude that the only starshades that would not be unfeasibly large would be near-UV/blue starshades (as opposed to NIR or full-bandpass starshades). Further in the future, the Remote Occulter\cite{Remote_Occulter} concept proposes an Earth-orbiting starshade operating with a ground-based extremely large telescope.

{\bf Morgan:} What are some open research questions that need to be worked on? How can I get involved?

{\bf Urania:} As starshades are a relatively new technology, there are many exciting research problems prime for exploration. First, there has not been much progress on the optimization of the apodization function for many years\cite{Vanderbei_2007, Cady_2010_b, Flamary_2014} and surely there is room for improvement; perhaps this could lead to the design of smaller, closer-in starshades.

Additionally, the numerically determined apodization functions are a black box of sorts and do not lend themselves to simple, analytic scaling relations. A clear set of scaling relations for numerically determined starshades would be greatly beneficial for those wanting to design a mission architecture for their science application.

Image processing techniques and the extraction of science from starshade images lag behind that of coronagraphic images and there is much to be done in this space (though progress has started in {\bf Hu, M. et al.} and {\bf Hu, R. et al. -- Data Challenge}). The SISTER software of {\bf Hildebrandt et al.} provides easy access to simulating starshade images and is a great place to start exploring the analysis of starshade observations.

There are too many other topics to list here, but I encourage you to read the papers in this issue to understand the current state of the field and for inspiration on where to take it next.

{\bf Morgan:} Where can I learn more about starshades?

{\bf Urania:} NASA's Exoplanet Program maintains a website\footnote{ \url{https://exoplanets.nasa.gov/exep/technology/starshade/}} with information about NASA's starshade technology development program. I have created a public ADS library\footnote{\url{https://ui.adsabs.harvard.edu/public-libraries/L5KELKzcTAq5lNfb9nynaQ}} including as many starshade publication as possible. I look forward to reading your contributions to the literature!

{\bf Morgan:} Thank you for taking the time to enlighten me on such a fascinating subject.

{\bf Urania:} My pleasure! I appreciate your interest in starshades; welcome to the community.

\subsection*{Disclosures}
The authors of this paper are members of the Starshade Science and Industry Partnership's Technology and Science Working Group. They have no relevant financial interests in the manuscript and declare no conflicts of interest.

\subsection* {Acknowledgments}
The authors would like to thank the many reviewers who provided their time and effort in carefully reviewing the papers in this special issue. We would also like to call out Gary Blackwood, Renyu Hu, and Phil Willems for their helpful comments on the editorial, and Sergi Hildebrandt for generating the simulated starshade images specifically for our Dialog.  Finally, a huge thank you to the editorial staff of JATIS, you have been helpful and supportive at every step of this journey.


\bibliography{references}   

\begin{thebibliography}{10}

\bibitem{Arenberg_Cash}
J.~W. {Arenberg} and W.~{Cash}, ``{New Worlds Observer: A Novel Mission Concept
  for Exoplanetary Studies},'' in {\em IAU Colloq. 200: Direct Imaging of
  Exoplanets: Science \& Techniques},  C.~{Aime} and F.~{Vakili}, Eds.,
  199--204  (2006).

\bibitem{Cash_2011}
W.~Cash, ``Analytic modeling of starshades,'' {\em ApJ} {\bf 738}, 76  (2011).

\bibitem{Cady_2012}
E.~{Cady}, ``{Boundary diffraction wave integrals for diffraction modeling of
  external occulters},'' {\em Optics Express} {\bf 20}, 15196  (2012).

\bibitem{Galileo}
G.~{Galilei}, {\em {Dialog Concerning the Two Chief World Systems}}, Modern
  Library, new edition~ed.  (2001).

\bibitem{Evans:48}
J.~W. Evans, ``A photometer for measurement of sky brightness near the sun,''
  {\em J. Opt. Soc. Am.} {\bf 38}, 1083--1085  (1948).

\bibitem{Newkirk:63}
G.~Newkirk and D.~Bohlin, ``Reduction of scattered light in the coronagraph,''
  {\em Appl. Opt.} {\bf 2}, 131--140  (1963).

\bibitem{Born_Wolf}
M.~{Born} and E.~{Wolf}, {\em {Principles of Optics}}, Cambridge University,
  7~ed.  (1999).

\bibitem{Boivin:78}
L.~P. Boivin, ``Reduction of diffraction errors in radiometry by means of
  toothed apertures,'' {\em Appl. Opt.} {\bf 17}, 3323--3328  (1978).

\bibitem{Shirley}
E.~Shirley and R.~Datla, ``Optimally toothed apertures for reduced
  diffraction,''  (1996).

\bibitem{Spitzer_1962}
J.~{Spitzer}, Lyman, ``{The Beginnings and Future of Space Astronomy},'' {\em
  American Scientist} {\bf 50}, 473--484  (1962).

\bibitem{Tousey_1965}
R.~{Tousey}, ``{Observations of the white light corona by rocket},'' {\em
  Annales d'Astrophysique} {\bf 28}, 600  (1965).

\bibitem{Harris_1978}
F.~Harris, ``On the use of windows for harmonic analysis with the discrete
  fourier transform,'' {\em Proceedings of the IEEE} {\bf 66}(1), 51--83
  (1978).

\bibitem{Marchal_1985}
C.~{Marchal}, ``{Concept of a space telescope able to see the planets and even
  the satellites around the nearest stars},'' {\em Acta Astronautica} {\bf 12},
  195--201  (1985).

\bibitem{Vanderbei_2007}
R.~J. Vanderbei, E.~Cady, and N.~J. Kasdin, ``Optimal occulter design for
  finding extrasolar planets,'' {\em ApJ} {\bf 665}, 794 -- 798  (2007).

\bibitem{Banderman}
L.~W. Bandermann, R.~H. Pohle, J.~P. Murphy, {\em et~al.}, ``Satellites for the
  detection of nonsolar planets,'' {\em Journal of Spacecraft and Rockets} {\bf
  18}(2), 164--171  (1981).

\bibitem{Cocks}
F.~H. Cocks, J.~E. Bischoff, S.~A. Watkins, {\em et~al.}, ``{Extrasolar
  planetary detection via stellar occultation: a novel "smaller, cheaper,
  faster" concept employing the Hubble Space Telescope},'' in {\em Space
  Telescopes and Instruments IV},  P.~Y. Bely and J.~B. Breckinridge, Eds.,
  {\bf 2807}, 74 -- 85, International Society for Optics and Photonics, SPIE
  (1996).

\bibitem{UMBRAS_99}
A.~B. Schultz, D.~J. Schroeder, I.~J. Jordan, {\em et~al.}, ``{Imaging planets
  about other stars with UMBRAS},'' in {\em Infrared Spaceborne Remote Sensing
  VII},  M.~Strojnik and B.~F. Andresen, Eds.,  {\bf 3759}, 49 -- 58,
  International Society for Optics and Photonics, SPIE  (1999).

\bibitem{BOSS}
C.~Copi, G.~Starkman, and L.~Lichodziejewski, ``The big occulting steerable
  satellite (boss),'' in {\em 2000 IEEE Aerospace Conference. Proceedings (Cat.
  No.00TH8484)},   {\bf 7}, 403--410 vol.7  (2000).

\bibitem{Cash_2006}
W.~Cash, ``{Detection of Earth-like planets around nearby stars using a
  petal-shaped occulter},'' {\em Nature} {\bf 442}, 51 -- 53  (2006).

\bibitem{Vanderbei_2003}
R.~J. {Vanderbei}, D.~N. {Spergel}, and N.~J. {Kasdin}, ``{Circularly Symmetric
  Apodization via Star-shaped Masks},'' {\em ApJ} {\bf 599}, 686--694  (2003).

\bibitem{SRM}
S.~{Seager}, N.~J. {Kasdin}, {\em et~al.}, ``{Starshade Rendezvous Probe},''
  (2019).
\newblock
  \url{https://smd-prod.s3.amazonaws.com/science-red/s3fs-public/atoms/files/Starshade2.pdf}
  Last accessed: 4/23/2021.

\bibitem{Arenberg_2007}
J.~W. {Arenberg}, A.~S. {Lo}, T.~M. {Glassman}, {\em et~al.}, ``{Optical
  performance of the New Worlds Occulter},'' {\em Comptes Rendus Physique} {\bf
  8}, 438--447  (2007).

\bibitem{Glassman_2009}
T.~{Glassman}, A.~S. {Lo}, J.~{Arenberg}, {\em et~al.}, ``{Starshade scaling
  relations},'' in {\em Techniques and Instrumentation for Detection of
  Exoplanets IV},  S.~B. {Shaklan}, Ed., {\em Society of Photo-Optical
  Instrumentation Engineers (SPIE) Conference Series} {\bf 7440}, 744013
  (2009).

\bibitem{Arenberg_2007_SPIE}
J.~{Arenberg}, T.~{Glassman}, and A.~S. {Lo}, ``{Effects of scattered light on
  the performance of the New Worlds Starshade},'' in {\em Techniques and
  Instrumentation for Detection of Exoplanets III},  D.~R. {Coulter}, Ed., {\em
  Society of Photo-Optical Instrumentation Engineers (SPIE) Conference Series}
  {\bf 6693}, 66931E  (2007).

\bibitem{Arenberg_2008_b}
J.~W. {Arenberg}, T.~{Glassman}, A.~S. {Lo}, {\em et~al.}, ``{New Worlds
  Observer system architecture},'' in {\em Space Telescopes and Instrumentation
  2008: Optical, Infrared, and Millimeter},  J.~{Oschmann}, Jacobus~M.,
  M.~W.~M. {de Graauw}, and H.~A. {MacEwen}, Eds., {\em Society of
  Photo-Optical Instrumentation Engineers (SPIE) Conference Series} {\bf 7010},
  70101S  (2008).

\bibitem{Keller_1962}
J.~B. {Keller}, ``{Geometrical theory of diffraction},'' {\em Journal of the
  Optical Society of America (1917-1983)} {\bf 52}, 116  (1962).

\bibitem{Savransky_2010}
D.~{Savransky}, N.~J. {Kasdin}, and E.~{Cady}, ``{Analyzing the Designs of
  Planet-Finding Missions},'' {\em PASP} {\bf 122}, 401  (2010).

\bibitem{Glassman_2011}
T.~{Glassman}, L.~{Newhart}, W.~{Voshell}, {\em et~al.}, ``{Creating optimal
  observing schedules for a starshade planet-finding mission},'' in {\em IEEE
  2011 Aerospace Conference},  E1  (2011).

\bibitem{Kolemen_2012}
E.~{Kolemen} and N.~J. {Kasdin}, ``{Optimization of an Occulter-Based
  Extrasolar-Planet-Imaging Mission},'' {\em Journal of Guidance Control
  Dynamics} {\bf 35}, 172--185  (2012).

\bibitem{Soto_2019}
G.~J. {Soto}, D.~{Savransky}, D.~{Garrett}, {\em et~al.}, ``{Parameterizing the
  Search Space of Starshade Fuel Costs for Optimal Observation Schedules},''
  {\em Journal of Guidance Control Dynamics} {\bf 42}, 2671--2676  (2019).

\bibitem{Hunyadi}
S.~L. Hunyadi, A.~S. Lo, and S.~B. Shaklan, ``{The dark side of TPF: detecting
  and characterizing extra-solar Earthlike planets with one or two external
  occulters},'' in {\em Techniques and Instrumentation for Detection of
  Exoplanets III},  D.~R. Coulter, Ed.,  {\bf 6693}, 26 -- 36, International
  Society for Optics and Photonics, SPIE  (2007).

\bibitem{Leviton_2007}
D.~B. {Leviton}, W.~C. {Cash}, B.~{Gleason}, {\em et~al.}, ``{White-light
  demonstration of one hundred parts per billion irradiance suppression in air
  by new starshade occulters},'' in {\em UV/Optical/IR Space Telescopes:
  Innovative Technologies and Concepts III},  H.~A. {MacEwen} and J.~B.
  {Breckinridge}, Eds., {\em Society of Photo-Optical Instrumentation Engineers
  (SPIE) Conference Series} {\bf 6687}, 66871B  (2007).

\bibitem{Schindhelm_2007}
E.~{Schindhelm}, A.~{Shipley}, P.~{Oakley}, {\em et~al.}, ``{Laboratory studies
  of petal-shaped occulters},'' in {\em Techniques and Instrumentation for
  Detection of Exoplanets III},  D.~R. {Coulter}, Ed., {\em Society of
  Photo-Optical Instrumentation Engineers (SPIE) Conference Series} {\bf 6693},
  669305  (2007).

\bibitem{Glassman_2013}
T.~{Glassman}, S.~{Casement}, S.~{Warwick}, {\em et~al.}, ``{Achieving
  high-contrast ratios with a 60-cm starshade},'' in {\em Techniques and
  Instrumentation for Detection of Exoplanets VI},  S.~{Shaklan}, Ed., {\em
  Society of Photo-Optical Instrumentation Engineers (SPIE) Conference Series}
  {\bf 8864}, 886418  (2013).

\bibitem{Glassman_2014}
T.~{Glassman}, S.~{Casement}, S.~{Warwick}, {\em et~al.}, ``{Measurements of
  high-contrast starshade performance},'' in {\em Space Telescopes and
  Instrumentation 2014: Optical, Infrared, and Millimeter Wave},
  J.~{Oschmann}, Jacobus~M., M.~{Clampin}, G.~G. {Fazio}, {\em et~al.}, Eds.,
  {\em Society of Photo-Optical Instrumentation Engineers (SPIE) Conference
  Series} {\bf 9143}, 91432O  (2014).

\bibitem{Smith_2016}
D.~{Smith}, S.~{Warwick}, T.~M. {Glassman}, {\em et~al.}, ``{Measurements of
  high-contrast starshade performance in the field},'' in {\em Space Telescopes
  and Instrumentation 2016: Optical, Infrared, and Millimeter Wave},  H.~A.
  {MacEwen}, G.~G. {Fazio}, M.~{Lystrup}, {\em et~al.}, Eds., {\em Society of
  Photo-Optical Instrumentation Engineers (SPIE) Conference Series} {\bf 9904},
  99043K  (2016).

\bibitem{Novicki_2016}
M.~C. {Novicki}, S.~{Warwick}, D.~K. {Smith}, {\em et~al.}, ``{Suppression of
  astronomical sources using the McMath-Pierce Solar Telescope and starshades
  with flight-like optics},'' in {\em Space Telescopes and Instrumentation
  2016: Optical, Infrared, and Millimeter Wave},  H.~A. {MacEwen}, G.~G.
  {Fazio}, M.~{Lystrup}, {\em et~al.}, Eds., {\em Society of Photo-Optical
  Instrumentation Engineers (SPIE) Conference Series} {\bf 9904}, 990426
  (2016).

\bibitem{Harness_2017}
A.~{Harness}, W.~{Cash}, and S.~{Warwick}, ``{High contrast observations of
  bright stars with a starshade},'' {\em Experimental Astronomy} {\bf 44},
  209--237  (2017).

\bibitem{Thomson_2010}
M.~W. Thomson, P.~D. Lisman, R.~Helms, {\em et~al.}, ``{Starshade design for
  occulter based exoplanet missions},'' in {\em Space Telescopes and
  Instrumentation 2010: Optical, Infrared, and Millimeter Wave},  J.~M.~O. Jr.,
  M.~C. Clampin, and H.~A. MacEwen, Eds.,  {\bf 7731}, 1757 -- 1769,
  International Society for Optics and Photonics, SPIE  (2010).

\bibitem{Lillie_2008}
C.~F. Lillie, D.~R. Dailey, A.~S. Lo, {\em et~al.}, ``{Deployment technologies
  for terrestrial planet finding missions},'' in {\em Advanced Optical and
  Mechanical Technologies in Telescopes and Instrumentation},  E.~Atad-Ettedgui
  and D.~Lemke, Eds.,  {\bf 7018}, 672 -- 683, International Society for Optics
  and Photonics, SPIE  (2008).

\bibitem{Kasdin_2011}
N.~J. {Kasdin}, D.~N. {Spergel}, R.~J. {Vanderbei}, {\em et~al.}, ``{Advancing
  technology for starlight suppression via an external occulter},'' in {\em
  Techniques and Instrumentation for Detection of Exoplanets V},  S.~{Shaklan},
  Ed., {\em Society of Photo-Optical Instrumentation Engineers (SPIE)
  Conference Series} {\bf 8151}, 81510J  (2011).

\bibitem{Webb_2014}
D.~{Webb}, N.~J. {Kasdin}, D.~{Lisman}, {\em et~al.}, ``{Successful Starshade
  petal deployment tolerance verification in support of NASA's technology
  development for exoplanet missions},'' in {\em Advances in Optical and
  Mechanical Technologies for Telescopes and Instrumentation},  R.~{Navarro},
  C.~R. {Cunningham}, and A.~A. {Barto}, Eds., {\em Society of Photo-Optical
  Instrumentation Engineers (SPIE) Conference Series} {\bf 9151}, 91511P
  (2014).

\bibitem{Webb_2019}
D.~Webb, M.~Arya, S.~Bradford, {\em et~al.}, ``{Advances in starshade
  technology readiness for an exoplanet characterizing science mission in the
  2020’s},'' in {\em Techniques and Instrumentation for Detection of
  Exoplanets IX},  S.~B. Shaklan, Ed.,  {\bf 11117}, 211 -- 225, International
  Society for Optics and Photonics, SPIE  (2019).

\bibitem{Webb_2020}
D.~Webb, S.~C. Bradford, J.~Steeves, {\em et~al.}, ``{Starshade Technology
  Development Activity {M}ilestone 8{A}: Verify Petal Position On-Orbit
  Stability},'' {\em Jet Propulsion Laboratory Publications}   (2020).
\newblock \url{https://exoplanets.nasa.gov/internal_resources/1696}.

\bibitem{Arya_2020}
M.~Arya, D.~Webb, F.~Mechentel, {\em et~al.}, ``{Starshade Technology
  Development Activity {M}ilestone 7{A}: Demonstration of Dimensional Stability
  of Perimeter Truss Bay Longeron and Node},'' {\em Jet Propulsion Laboratory
  Publications}   (2020).
\newblock \url{https://exoplanets.nasa.gov/internal_resources/1677}.

\bibitem{Shaklan_2010}
S.~B. {Shaklan}, M.~C. {Noecker}, T.~{Glassman}, {\em et~al.}, ``{Error
  budgeting and tolerancing of starshades for exoplanet detection},'' in {\em
  Space Telescopes and Instrumentation 2010: Optical, Infrared, and Millimeter
  Wave},  J.~{Oschmann}, Jacobus~M., M.~C. {Clampin}, and H.~A. {MacEwen},
  Eds., {\em Society of Photo-Optical Instrumentation Engineers (SPIE)
  Conference Series} {\bf 7731}, 77312G  (2010).

\bibitem{Shaklan_2015}
S.~B. {Shaklan}, L.~{Marchen}, E.~{Cady}, {\em et~al.}, ``{Error budgets for
  the Exoplanet Starshade (Exo-S) probe-class mission study},'' in {\em
  Techniques and Instrumentation for Detection of Exoplanets VII},
  S.~{Shaklan}, Ed., {\em Society of Photo-Optical Instrumentation Engineers
  (SPIE) Conference Series} {\bf 9605}, 96050Z  (2015).

\bibitem{Glassman_2016}
T.~{Glassman}, S.~{Warwick}, A.~{Lo}, {\em et~al.}, ``{Starshade
  starlight-suppression performance with a deployable structure},'' in {\em
  Space Telescopes and Instrumentation 2016: Optical, Infrared, and Millimeter
  Wave},  H.~A. {MacEwen}, G.~G. {Fazio}, M.~{Lystrup}, {\em et~al.}, Eds.,
  {\em Society of Photo-Optical Instrumentation Engineers (SPIE) Conference
  Series} {\bf 9904}, 990425  (2016).

\bibitem{Flinois_2020}
T.~L.~B. {Flinois}, D.~P. {Scharf}, C.~R. {Seubert}, {\em et~al.}, ``{Starshade
  formation flying II: formation control},'' {\em Journal of Astronomical
  Telescopes, Instruments, and Systems} {\bf 6}, 029001  (2020).

\bibitem{Noecker_2007}
M.~C. {Noecker}, ``{Alignment of a terrestrial planet finder starshade at
  20-100 megameters},'' in {\em Techniques and Instrumentation for Detection of
  Exoplanets III},  D.~R. {Coulter}, Ed., {\em Society of Photo-Optical
  Instrumentation Engineers (SPIE) Conference Series} {\bf 6693}, 669306
  (2007).

\bibitem{Leitner_2007}
J.~{Leitner}, ``{Formation flying system design for a planet-finding
  telescope-occulter system},'' in {\em UV/Optical/IR Space Telescopes:
  Innovative Technologies and Concepts III},  H.~A. {MacEwen} and J.~B.
  {Breckinridge}, Eds., {\em Society of Photo-Optical Instrumentation Engineers
  (SPIE) Conference Series} {\bf 6687}, 66871D  (2007).

\bibitem{Martin_2015}
S.~{Martin}, D.~{Scharf}, E.~{Cady}, {\em et~al.}, ``{Optical instrumentation
  for science and formation flying with a starshade observatory},'' in {\em
  Techniques and Instrumentation for Detection of Exoplanets VII},
  S.~{Shaklan}, Ed., {\em Society of Photo-Optical Instrumentation Engineers
  (SPIE) Conference Series} {\bf 9605}, 96050X  (2015).

\bibitem{Scharf_2016}
D.~P. {Scharf}, S.~R. {Martin}, C.~C. {Liebe}, {\em et~al.}, ``{Precision
  formation flying at megameter separations for exoplanet characterization},''
  {\em Acta Astronautica} {\bf 123}, 420--434  (2016).

\bibitem{Flinois_2018_S5}
T.~Flinois, M.~Bottom, S.~Martin, {\em et~al.}, ``{Starshade Technology
  Development Activity {M}ilestone 4: Lateral Formation Sensing and Control},''
  {\em Jet Propulsion Laboratory Publications}   (2018).
\newblock \url{https://exoplanets.nasa.gov/internal_resources/1077}.

\bibitem{Koenig_2019}
A.~W. {Koenig}, B.~{Macintosh}, and S.~{D'Amico}, ``{Formation Design of
  Distributed Telescopes in Earth Orbit for Astrophysics Applications},'' {\em
  Journal of Spacecraft and Rockets} {\bf 56}, 1462--1477  (2019).

\bibitem{Bottom_2020}
M.~{Bottom}, S.~{Martin}, E.~{Cady}, {\em et~al.}, ``{Starshade formation
  flying I: optical sensing},'' {\em Journal of Astronomical Telescopes,
  Instruments, and Systems} {\bf 6}, 015003  (2020).

\bibitem{Palacios_2020}
L.~M. {Palacios}, A.~{Harness}, and N.~J. {Kasdin}, ``{Hardware-in-the-loop
  testing of formation flying control and sensing algorithms for starshade
  missions},'' {\em Acta Astronautica} {\bf 171}, 97--105  (2020).

\bibitem{Harness_2018}
A.~{Harness}, S.~{Shaklan}, W.~{Cash}, {\em et~al.}, ``{Advances in edge
  diffraction algorithms},'' {\em Journal of the Optical Society of America A}
  {\bf 35}, 275  (2018).

\bibitem{Harness_2020}
A.~{Harness}, ``{Implementing non-scalar diffraction in Fourier optics via the
  Braunbek method},'' {\em Optics Express} {\bf 28}, 34290  (2020).

\bibitem{Rocco_2009}
R.~{Samuele}, T.~{Glassman}, A.~M.~J. {Johnson}, {\em et~al.}, ``{Starlight
  suppression from the starshade testbed at NGAS},'' in {\em Techniques and
  Instrumentation for Detection of Exoplanets IV},  S.~B. {Shaklan}, Ed., {\em
  Society of Photo-Optical Instrumentation Engineers (SPIE) Conference Series}
  {\bf 7440}, 744004  (2009).

\bibitem{Rocco_2010}
R.~{Samuele}, R.~{Varshneya}, T.~P. {Johnson}, {\em et~al.}, ``{Progress at the
  starshade testbed at Northrop Grumman Aerospace Systems: comparisons with
  computer simulations},'' in {\em Space Telescopes and Instrumentation 2010:
  Optical, Infrared, and Millimeter Wave},  J.~{Oschmann}, Jacobus~M., M.~C.
  {Clampin}, and H.~A. {MacEwen}, Eds., {\em Society of Photo-Optical
  Instrumentation Engineers (SPIE) Conference Series} {\bf 7731}, 773151
  (2010).

\bibitem{Cady_2010}
E.~{Cady}, K.~{Balasubramanian}, M.~{Carr}, {\em et~al.}, ``{Broadband
  suppression and occulter position sensing at the Princeton occulter
  testbed},'' in {\em Space Telescopes and Instrumentation 2010: Optical,
  Infrared, and Millimeter Wave},  J.~{Oschmann}, Jacobus~M., M.~C. {Clampin},
  and H.~A. {MacEwen}, Eds., {\em Society of Photo-Optical Instrumentation
  Engineers (SPIE) Conference Series} {\bf 7731}, 77312F  (2010).

\bibitem{Sirbu_2013}
D.~{Sirbu}, N.~J. {Kasdin}, and R.~J. {Vanderbei}, ``{Monochromatic
  verification of high-contrast imaging with an occulter},'' {\em Optics
  Express} {\bf 21}, 32234  (2013).

\bibitem{Harness_2021}
A.~Harness, S.~Shaklan, P.~Willems, {\em et~al.}, ``{Optical verification
  experiments of sub-scale starshades},'' {\em Journal of Astronomical
  Telescopes, Instruments, and Systems} {\bf 7}(2), 1 -- 32  (2021).

\bibitem{Hu_2021_noise}
R.~Hu, D.~Lisman, S.~Shaklan, {\em et~al.}, ``{Overview and reassessment of
  noise budget of starshade exoplanet imaging},'' {\em Journal of Astronomical
  Telescopes, Instruments, and Systems} {\bf 7}(2), 1 -- 24  (2021).

\bibitem{Shaklan_2021}
S.~Shaklan, E.~Hilgemann, D.~McKeithen, {\em et~al.}, ``{Solar glint from
  uncoated starshade optical edges},'' {\em Journal of Astronomical Telescopes,
  Instruments, and Systems} {\bf 7}(2), 1 -- 19  (2021).

\bibitem{Casement_2012}
S.~{Casement}, M.~{Flannery}, T.~{Glassman}, {\em et~al.}, ``{Starshade design
  driven by stray light from edge scatter},'' in {\em Space Telescopes and
  Instrumentation 2012: Optical, Infrared, and Millimeter Wave},  M.~C.
  {Clampin}, G.~G. {Fazio}, H.~A. {MacEwen}, {\em et~al.}, Eds., {\em Society
  of Photo-Optical Instrumentation Engineers (SPIE) Conference Series} {\bf
  8442}, 84424H  (2012).

\bibitem{Martin_2013}
S.~R. {Martin}, S.~{Shaklan}, S.~{Crawford}, {\em et~al.}, ``{Starshade optical
  edge modeling, requirements, and laboratory tests},'' in {\em Techniques and
  Instrumentation for Detection of Exoplanets VI},  S.~{Shaklan}, Ed., {\em
  Society of Photo-Optical Instrumentation Engineers (SPIE) Conference Series}
  {\bf 8864}, 88641A  (2013).

\bibitem{Casement_2016}
S.~{Casement}, S.~{Warwick}, D.~{Smith}, {\em et~al.}, ``{Results of edge
  scatter testing for a starshade mission},'' in {\em Space Telescopes and
  Instrumentation 2016: Optical, Infrared, and Millimeter Wave},  H.~A.
  {MacEwen}, G.~G. {Fazio}, M.~{Lystrup}, {\em et~al.}, Eds., {\em Society of
  Photo-Optical Instrumentation Engineers (SPIE) Conference Series} {\bf 9904},
  99043H  (2016).

\bibitem{Steeves_2016}
J.~{Steeves}, S.~{Martin}, D.~{Webb}, {\em et~al.}, ``{Precision optical edges
  for a starshade external occulter},'' in {\em Advances in Optical and
  Mechanical Technologies for Telescopes and Instrumentation II},  R.~{Navarro}
  and J.~H. {Burge}, Eds., {\em Society of Photo-Optical Instrumentation
  Engineers (SPIE) Conference Series} {\bf 9912}, 99122O  (2016).

\bibitem{Steeves_2018}
J.~{Steeves}, H.~J. {Lee}, E.~{Hilgemann}, {\em et~al.}, ``{Development of
  low-scatter optical edges for starshades},'' in {\em Advances in Optical and
  Mechanical Technologies for Telescopes and Instrumentation III},
  R.~{Navarro} and R.~{Geyl}, Eds., {\em Society of Photo-Optical
  Instrumentation Engineers (SPIE) Conference Series} {\bf 10706}, 107065K
  (2018).

\bibitem{Hilgemann_2019}
E.~{Hilgemann}, D.~{McKeithen}, N.~{Saltarelli}, {\em et~al.}, ``{Advancements
  in precision edges for a starshade external occulter},'' in {\em Society of
  Photo-Optical Instrumentation Engineers (SPIE) Conference Series},  {\em
  Society of Photo-Optical Instrumentation Engineers (SPIE) Conference Series}
  {\bf 11117}, 111170Q  (2019).

\bibitem{Hilgemann_2019_S5}
E.~Hilgemann, S.~Shaklan, D.~McKeithen, {\em et~al.}, ``{Starshade Technology
  Development Activity {M}ilestone 3: Demonstration of Solar Glint Lobe Scatter
  Performance},'' {\em Jet Propulsion Laboratory Publications}   (2019).
\newblock \url{https://exoplanets.nasa.gov/internal_resources/1544}.

\bibitem{McKeithen_2019}
D.~{McKeithen}, S.~{Shaklan}, S.~{Martin}, {\em et~al.}, ``{Modeling the
  scatter of sunlight from starshade edges},'' in {\em Society of Photo-Optical
  Instrumentation Engineers (SPIE) Conference Series},  {\em Society of
  Photo-Optical Instrumentation Engineers (SPIE) Conference Series} {\bf
  11117}, 111171L  (2019).

\bibitem{HabEx_Report}
B.~S. {Gaudi}, S.~{Seager}, B.~{Mennesson}, {\em et~al.}, ``{The Habitable
  Exoplanet Observatory (HabEx) Mission Concept Study Final Report},'' {\em
  arXiv e-prints} , arXiv:2001.06683  (2020).

\bibitem{Shipley_2007}
A.~{Shipley}, W.~{Cash}, J.~W. {Arenberg}, {\em et~al.}, ``{New Worlds Observer
  tolerance overview},'' in {\em UV/Optical/IR Space Telescopes: Innovative
  Technologies and Concepts III},  H.~A. {MacEwen} and J.~B. {Breckinridge},
  Eds., {\em Society of Photo-Optical Instrumentation Engineers (SPIE)
  Conference Series} {\bf 6687}, 66871A  (2007).

\bibitem{Arenberg_2008}
J.~W. {Arenberg}, A.~{Shipley}, W.~{Cash}, {\em et~al.}, ``{Sensitivity
  analysis of the New Worlds starshade's shadow},'' in {\em Space Telescopes
  and Instrumentation 2008: Optical, Infrared, and Millimeter},  J.~{Oschmann},
  Jacobus~M., M.~W.~M. {de Graauw}, and H.~A. {MacEwen}, Eds., {\em Society of
  Photo-Optical Instrumentation Engineers (SPIE) Conference Series} {\bf 7010},
  70101V  (2008).

\bibitem{Shaklan_2011}
S.~B. {Shaklan}, L.~{Marchen}, P.~D. {Lisman}, {\em et~al.}, ``{A starshade
  petal error budget for exo-earth detection and characterization},'' in {\em
  Techniques and Instrumentation for Detection of Exoplanets V},  S.~{Shaklan},
  Ed., {\em Society of Photo-Optical Instrumentation Engineers (SPIE)
  Conference Series} {\bf 8151}, 815113  (2011).

\bibitem{Shaklan_2017}
S.~B. {Shaklan}, L.~{Marchen}, and E.~{Cady}, ``{Shape accuracy requirements on
  starshades for large and small apertures},'' in {\em Society of Photo-Optical
  Instrumentation Engineers (SPIE) Conference Series},  {\em Society of
  Photo-Optical Instrumentation Engineers (SPIE) Conference Series} {\bf
  10400}, 104001T  (2017).

\bibitem{Kane_2020}
S.~R. {Kane}, M.~C. {Turnbull}, B.~J. {Fulton}, {\em et~al.}, ``{Dynamical
  Packing in the Habitable Zone: The Case of Beta CVn},'' {\em AJ} {\bf 160},
  81  (2020).

\bibitem{Romero-Wolf_2021}
A.~Romero-Wolf, G.~Bryden, S.~Seager, {\em et~al.}, ``{Starshade rendezvous:
  exoplanet sensitivity and observing strategy},'' {\em Journal of Astronomical
  Telescopes, Instruments, and Systems} {\bf 7}(2), 1 -- 28  (2021).

\bibitem{Roman_detector}
``{Nancy Grace Roman Space Telescope Simulations - Spacecraft and Instrument
  Parameters},''  (2021).
\newblock \url{https://wfirst.ipac.caltech.edu/sims/Param_db.html} Last
  accessed: 4/30/2021.

\bibitem{Stark_2016}
C.~C. {Stark}, E.~J. {Cady}, M.~{Clampin}, {\em et~al.}, ``{A direct comparison
  of exoEarth yields for starshades and coronagraphs},'' in {\em Space
  Telescopes and Instrumentation 2016: Optical, Infrared, and Millimeter Wave},
   H.~A. {MacEwen}, G.~G. {Fazio}, M.~{Lystrup}, {\em et~al.}, Eds., {\em
  Society of Photo-Optical Instrumentation Engineers (SPIE) Conference Series}
  {\bf 9904}, 99041U  (2016).

\bibitem{DAmico_2019}
S.~{D'Amico}, A.~{Koenig}, B.~{Macintosh}, {\em et~al.}, ``{System Design of
  the Miniaturized Distributed Occulter/Telescope (mDOT) Science Mission},'' in
  {\em Proceedings of the AIAA/USU Conference on Small Satellites},  81
  (2019).

\bibitem{Remote_Occulter}
E.~{Peretz} and J.~{Mather}, ``{Remote Occulter: An orbiting starshade working
  with Extremely large telescopes on the ground to study exoplanets and
  planetary systems},''  (2020).
\newblock \url{https://exoplanets.nasa.gov/internal_resources/1538/} Last
  accessed: 5/10/2021.

\bibitem{Peretz_2019}
E.~{Peretz}, J.~{Mather}, S.~{Seager}, {\em et~al.}, ``{Mapping the observable
  sky for a remote occulter working with ground-based telescopes},'' in {\em
  Society of Photo-Optical Instrumentation Engineers (SPIE) Conference Series},
   {\em Society of Photo-Optical Instrumentation Engineers (SPIE) Conference
  Series} {\bf 11117}, 111170S  (2019).

\bibitem{SSWG_charter}
N.~E. Office, ``{StarShade Readiness Working Group (SSWG) - Charter},''
  (2016).
\newblock
  \url{https://exoplanets.nasa.gov/exep/files/exep/SSWG\%20Charter_finalfinal.pdf}
  Last accessed: 4/23/2021.

\bibitem{SSWG_briefing}
``{Starshade Readiness Working Group Recommendation to Astrophysics Division
  Director},''  (2016).
\newblock
  \url{https://exoplanets.nasa.gov/system/internal_resources/details/original/339_SSWG_APD_briefing_final.pdf}
  Last accessed: 4/23/2021.

\bibitem{LUVOIR_Report}
{The LUVOIR Team}, ``{The LUVOIR Mission Concept Study Final Report},'' {\em
  arXiv e-prints} , arXiv:1912.06219  (2019).

\bibitem{Redding_2019}
D.~{Redding}, K.~{Coste}, O.~{Polanco}, {\em et~al.}, ``{A Habitable Exoplanet
  Observatory (HabEx) starshade-only architectures},'' in {\em UV/Optical/IR
  Space Telescopes and Instruments: Innovative Technologies and Concepts IX},
  {\em Society of Photo-Optical Instrumentation Engineers (SPIE) Conference
  Series} {\bf 11115}, 111150V  (2019).

\bibitem{Soummer_2009}
R.~{Soummer}, W.~{Cash}, R.~A. {Brown}, {\em et~al.}, ``{A starshade for JWST:
  science goals and optimization},'' in {\em Techniques and Instrumentation for
  Detection of Exoplanets IV},  S.~B. {Shaklan}, Ed., {\em Society of
  Photo-Optical Instrumentation Engineers (SPIE) Conference Series} {\bf 7440},
  74400A  (2009).

\bibitem{Cady_2010_b}
E.~{Cady}, N.~J. {Kasdin}, and S.~{Shaklan}, ``{Designing asymmetric and
  branched petals for planet-finding occulters},'' {\em Optics Express} {\bf
  18}, 523  (2010).

\bibitem{Flamary_2014}
R.~{Flamary} and C.~{Aime}, ``{Optimization of starshades: focal plane versus
  pupil plane},'' {\em Astronomy \& Astrophysics} {\bf 569}, A28  (2014).

\end{thebibliography}
\bibliographystyle{spiejour}   

{\bf Author Biographies}

\vspace{2ex}\noindent\textbf{Jonathan Arenberg} is the Chief Mission Architect for Science and Robotic Exploration  at Northrop Grumman Space Systems. His degrees are in physics (BS '83) and engineering (MS '85, Ph.D. '87) from the University of California, Los Angeles.  He has worked on the Chandra X-ray Observatory, the Starshade and James Webb Space Telescope and many mission studies.

\vspace{2ex}\noindent\textbf{Anthony Harness} is an Associate Research Scholar in the Mechanical and Aerospace Engineering Department at Princeton University. He received his Ph.D. in Astrophysics in 2016 from the University of Colorado Boulder. He currently leads the experiments at Princeton validating starshade optical technologies.

\vspace{2ex}\noindent\textbf{Rebecca Jensen-Clem} is an Assistant Professor of Astronomy \& Astrophysics at the University of California Santa Cruz. She received her Ph.D. in Astrophysics in 2017 from Caltech. Her research focuses on extreme adaptive optics technology development and observational exoplanet science.


\end{spacing}
\end{document}